\documentclass{article}
\usepackage{amsmath}
\usepackage{graphicx}
\usepackage{nopageno}
\usepackage{amssymb}
\usepackage{hyperref}
\usepackage[toc,page]{appendix}
\usepackage[margin=1.0 in]{geometry}
\usepackage[center]{titlesec}

\hypersetup{linktocpage}
\newtheorem{theorem}{Theorem}
\newtheorem{acknowledgement}[theorem]{Acknowledgement}

\newtheorem{remark}[theorem]{Remark}

\input{tcilatex}
\renewenvironment{abstract}
 {\par\noindent\textbf{\abstractname.}\ \ignorespaces}
 {\par\medskip}
\newlength{\bibitemsep}
\newlength{\bibparskip}
\let\oldthebibliography\thebibliography
\renewcommand\thebibliography[1]{  \oldthebibliography{#1}  \setlength{\parskip}{\bibitemsep}  \setlength{\itemsep}{\bibparskip}}

\begin{document}

\title{\textbf{On quantum Hall effect, Kosterlitz-Thouless phase transition,
Dirac magnetic monopole, and Bohr-Sommerfeld quantization }}
\author{Felix A. Buot$^{1,2}$, Allan Roy Elnar$^{1,3}$, Gibson Maglasang$%
^{1,3}$, and Roland E.S. Otadoy$^{1}$ \\
$^{1}$\textit{Laboratory of Computational Functional Materials, }\\
\textit{Nanoscience and Nanotechnology (LCFMNN), }\\
\textit{University of San Carlos, Talamban, Cebu City 6000, Philippines,}\\
$^{2}$\textit{C\&LB Research Institute, Carmen, Cebu 6005, Philippines,}\\
$^{3}$\textit{Cebu Normal University, Cebu City, 6000, Philippines}\\
\textit{\ }}
\date{}
\maketitle

\begin{abstract}
We addressed quantization phenomena in transport and
vortex/precession-motion of low-dimensional systems, stationary quantization
of confined motion in phase space due to oscillatory dynamics or
compactification of space and time for steady-state systems (e.g., particle
in a box or torus, Brillouin zone, and Matsubara time zone or Matsubara
quantized frequencies), and the quantization of sources. We discuss how the
self-consistent Bohr-Sommerfeld quantization condition permeates the
relationships between the quantization of integer Hall effect, fractional
quantum Hall effect, the Berezenskii-Kosterlitz-Thouless vortex
quantization, the Dirac magnetic monopole, the Haldane phase, contact
resistance in closed mesoscopic circuits of quantum physics, and in the
monodromy (holonomy) of completely integrable Hamiltonian systems of quantum
geometry. In quantum transport of open systems, quantization occurs in
fundamental units of quantum conductance, other closed systems in quantum
units dictated by Planck's constant, and for sources in units of discrete
vortex charge and Dirac magnetic monopole charge. The thesis of the paper is
that if we simply cast the B-S quantization condition as a U(1) gauge
theory, like the gauge field of the topological quantum field theory (TQFT)
via the Chern-Simons gauge theory, or specifically as in topological band
theory (TBT) of condensed matter physics in terms of Berry connection and
curvature to make it self-consistent, then all the quantization method in
all the physical phenomena treated in this paper are unified. This paper is
motivated by the recent derivation by one of the authors of the integer
quantum Hall effect in electrical conductivity using a novel phase-space
nonequilibrium quantum transport approach. All of the above quantization of
physical phenomena may thus be unified simply from the geometric point of
view of the old Bohr-Sommerfeld quantization, as a theory of Berry
connection in parallel transport or as a U(1) gauge theory.
\end{abstract}

\tableofcontents

\renewcommand{\refname}{REFERENCES}

\makeatletter\renewcommand\@biblabel[1]{#1.} \makeatother

\section{Introduction}

There are two general themes that will be addressed in this paper, namely,
(a) transport in open systems and (b) vortex/precession-motion quantization
of low-dimensional systems, and stationary quantization of \textit{confined
motion} in phase space due to oscillatory dynamics or \textit{%
compactification} of space and time for steady-state systems (e.g., particle
in a box, Brillouin zone, and Matsubara time zone or Matsubara quantized
frequencies). The former is a recent phenomenon which has become the
springboard of modern condensed-matter physics research, whereas the latter
is as old as the beginning of quantum mechanics of atomic systems. Although,
the latter is old it continues to be interesting and a powerful way of
getting a handle of complex problems, and indeed it has now merited revisits
since its innocent looking formalism may actually be of rigorous geometric
origin. This is based on the geometrical concept of connection \cite{berry}.
Our task is to show that the old is the forebear of the new and permeates or
pervades the whole of modern topological quantum physics. The geometrical
concept of connection also allows us to deduce the quantized Dirac magnetic
monopole without the use of Dirac semi-infinite string \cite{dirac}.

The quantization of Hall conductance of electrical conductivity in a
two-dimensional periodic potential was first explained by Thouless, Kohmoto,
Nightingale, and den Nijs (TKNN)\cite{tknn} using the Kubo current-current
correlation. A similar approach was employed by Streda \cite{streda}.
Earlier, Laughlin\cite{laughlin}, and later Halperin \cite{halperin}, study
the effects produced by changes in the vector potential on the states at the
edges of a finite system, where quantization of the conductance is made
explicit, but whether their result is insensitive to boundary conditions was
not clear. In contrast, the use of Kubo formula by TKNN is for bulk
two-dimensional conductors.

These theoretical works were motivated by the 1980 Nobel Prize winning
experimental discovery of von Klitzing, Dorda, and Pepper\cite{kdp} on the
quantization of the Hall conductance of a two-dimensional electron gas in a
strong magnetic field. The strong magnetic field basically provides the
gapped energy structure for the experiments of two-dimensional electron gas.
In the TKNN approach, periodic potential in crystalline solid is being
treated. A strong magnetic field is not needed to provide the gapped energy
structure in their theory, only peculiar, gapped energy-band structures. In
principle, in the presence of electric field the discrete Landau levels is
replaced by the \textit{unstable} discrete Stark ladder-energy levels in the
absence of scatterings \cite{zener, wannier} due to compactification of
space into a torus, known as the Born-von Karman boundary condition.

Our new approach to integer quantum Hall effect (IQHE) makes use of the
real-time superfield and lattice Weyl transform nonequilibrium Green's
function (SFLWT-NEGF) \cite{buotbook} quantum transport equation. This is
given to first order in the gradient expansion \cite{trHn, bj} of the
quantum transport equation. This purely nonequilibrium quantum transport
approach is in contrast to the use of conventional equilibrium-fluctuation
Kubo formula originally employed by TKNN \cite{tknn}. The topological
invariant of quantum transport in $\left( \vec{p},\vec{q};E,t\right) $-space
is thus identified. How the B-S quantization enters into the analysis is
here identified. For a self-contained treatment, we give the derivation in
the text and in the Appendices. The Kubo current-current formula is also
derived in Appendix C from the SFLWT-NEGF quantum transport. The B-S
quantization also enters in the calculation leading to quantized orbital
magnetic moment and edge states.

Another Nobel prize winning theoretical discovery \cite{tknn}, and confirmed
experimentally, is the so-called Kosterlitz-Thouless phase transition (KTPT)
first discovered in $X$-$Y$ model of spin systems. This goes well-beyond the
well-established theorem, the so-called Mermin-Wagner theorem \cite{mermin},
sometimes referred to as the Mermin--Wagner--Berezinskii theorem, which
states that continuous symmetries cannot be spontaneously broken at finite
temperature in systems with sufficiently short-range interactions in
dimensions $d\leq 2$. The vortex and antivortex solutions of KTPT goes
beyond Ginzburg-Landau symmetry breaking phase transition which been quite
successful in the past. On the other hand, the vortex solutions of KTPT has
a topological content (basically topological defects) and bears resemblance
to the B-S quantization condition. The B-S quantization also crept into the
theory of another novel topological phase of matter, so called Haldane phase
of odd-integer spin chain. The emergence of new topological phases of matter
which violates symmetry breaking has ushered the interaction of theoretical
physics and pure mathematics.

In recent experiment reveals the appearance of incompressible fractional
quantum Hall effect (FQHE) states in monolayer graphene at $\nu =\pm \frac{1%
}{2}$,$\pm \frac{1}{4}$ and $\pm \frac{3}{4}$ substituting the compressible
Hall metal states at these fillings in the lowest Landau level in a narrow
magnetic field window depending on the sample parameters. Jacak propose \cite%
{jacak2} an explanation of these observations in terms of homotopy of
monolayer graphene in consistence with a general theory of correlated states
in planar Hall systems, which is beyond the composite fermion model of Jain 
\cite{jain}. An increase of the magnetic field flux quantum is proven for
multiloop trajectories (the formal proof goes via the Bohr--Sommerfeld rule 
\cite{jacak1}), resulting in the magnetic flux quantum, $\Phi _{k}=(2k+1)%
\frac{h}{e}$ , for planar braids with additional $k$ loops \cite{jacak2}.
More recently the B-S quantization has been employed by Jacak \cite{jacak3}
to show that a larger flux quantum for multi-loop cyclotron braid orbits
defines larger spatial dimension of such orbits. Generalization of the
Bohr-Sommerfeld quantization rule shows that the size of a magnetic field
flux quantum grows for multiloop orbits like $\left( 2k+1\right) $ $\frac{h}{%
c}$ with the number of loops $k$. Utilizing this property for electrons on
the 2-D substrate jellium, the author has derived upon the path integration
a complete FQHE hierarchy in excellent agreementt with experiments \cite%
{jacak2}.

Indeed, the recent developments in the cross-fertilization of pure
mathematics and physics is invigorated due to recent discoveries in physics
which have strong relevance to pure mathematics. The IQHE and fractional
FQHE have recently been geometrically formalized and unified based on Hecke
operators and Hecke eigensheaves (mathematical replacement of the term
eigenfunctions) i.e., of the geometric Langlands program by Ikeda \cite%
{ikeda}. In particular the plateaus of the Hall conductance are associated
by the Hecke eigensheaves. Moreover, the fractal energy spectrum of a
tight-binding Hamiltonian in a magnetic field, known as the Hofstadter's
butterfly, as been associated with Langlands duality in the quantum groups.
Indeed, the Langlands program in pure mathematics may carry several
realizations in theoretical physics.

As it turns out, the old Bohr-Sommerfeld quantization rules has recently
been realized to have much deeper ramifications in quantum physics. In a
much simpler terms in some cases, it has to do with the counting of discrete
"$\left( 2\pi \hbar \right) ^{3}$-voxels" of actions in phase-space volume
of generalized canonical variables measured in units of Planck's constant, $%
h,$\ or number of quantum flux in the case of magnetic fields. Contour
integral of Berry connection are measured in terms of $2\pi \hbar $-pixels.
By counting operation means that the results belong to the domain of
integers, $%
\mathbb{Z}
$.

Here we show that the old quantization scheme have bearings on Landau-level
degeneracy, massive Dirac magnetic monopole, Aharonov-Bohm effect, discrete
vortex charge, quantization of orbital magnetic moment and angular momentum,
and Bohr-Sommerfeld quantization in superfluidity, superconductivity, and
topological band theory (TBT) and topological quantum field theory (TQFT).

\section{Geometric origin of the Bohr-Sommerfeld quantization}

\bigskip We give here a geometrical point of view of the Bohr-Sommerfeld
(B-S) quantization rule given by,

\begin{equation}
\frac{1}{2\pi \hbar }\oint\limits_{H(p,q)=E}p_{i}dq_{i}=\frac{1}{2\pi \hbar }%
\oint\limits_{H(p,q)=E}\vec{p}\cdot d\vec{q}=n\text{.}  \label{b-s_eq}
\end{equation}%
The energy spectrum of hydrogen atom agreed exactly with observed spectrum
as obtained by Sommerfeld. \ The B-S theory was applied with varying success
to other systems. In fact, it is remarkable that the energy spectrum of
Dirac equation for an electron in an atom obtained by Sommerfeld also agrees
exactly \cite{davydov}. With a minor modifications B-S quantization gives
the energy spectrum of $\pi $ meson, which can be obtained by solving the
Klein-Gordon equation \cite{sniatycki}.

Closer examination of Eq. (\ref{b-s_eq}) suggests that it does not really
make any sense if one considers the set $\left\{ p,q\right\} $ as
independent conjugate dynamical variables. Obviously, it is totally
inconsistent with the surface integral%
\begin{equation}
\frac{1}{2\pi \hbar }\oint\limits_{H(p,q)=E\ \left[ \partial S\right] }\vec{p%
}\cdot d\vec{q}=\frac{1}{2\pi \hbar }\iint\limits_{S}\nabla _{\vec{q}}\times 
\vec{p}\cdot d\vec{S}=0\text{.}  \label{inconist}
\end{equation}%
The only pausible quantity that could be applied to give a new meaning to
the variable $\vec{p}$ in Eq. (\ref{b-s_eq}) is the vector potential $%
A\left( q\right) $ in Maxwell's equations. Then the integer $n$ would give
the Landau level degeneracies or the number of quantized vortices in type II
superconductors. But in the absence of magnetic fields, in which the B-S is
often applied, we have to appeal to more general gauge theories, e.g., the
gauge of TQFT via the Chern-Simons gauge theory \cite{wu} or the gauge of
TBT in terms of Berry connection and Berry curvature.

Here we give a much simpler and intuitive geometric derivation of the B-S
quantization condition, as a contour integral of the gradient of geometric
phase or Berry connection. Let $\left\vert n,q\right\rangle $ be the basis
eigenvector of the eigenstates of the Schr\"{o}dinger equation, parametrized
by $q$,%
\begin{equation*}
H\left( q\right) \left\vert n,q\right\rangle =E_{n}\left\vert
n,q\right\rangle .
\end{equation*}%
Note from the original quantum mechanical viewpoint, $p_{i}$ and $dq_{i}$ in
B-S quantization are $c$-numbers not operators. These $c$-numbers must
actually be derived from the Schr\"{o}dinger wavefunctions, adiabatically,
i.e., as diagonal matrix element of the momentum operator, $i\hbar \nabla
_{q}$ when operating on the eigenvector $\left\vert n,q\right\rangle $%
\begin{eqnarray*}
\frac{1}{2\pi \hbar }\oint p_{i}dq_{i} &=&\frac{1}{2\pi \hbar }\oint
\left\langle n,q\right\vert \left( i\hbar \nabla _{q_{i}}\right) \left\vert
n,q\right\rangle dq_{i} \\
&=&\frac{1}{2\pi \hbar }\left( i\hbar \right) \oint \left\langle
n,q\right\vert \nabla _{q_{i}}\left\vert n,q\right\rangle dq_{i}\text{.}
\end{eqnarray*}%
Thus,%
\begin{eqnarray*}
\frac{1}{2\pi \hbar }\oint p_{i}dq_{i} &=&\frac{i}{2\pi }\oint \left\langle
n,q\right\vert \nabla _{q_{i}}\left\vert n,q\right\rangle dq_{i} \\
&=&\frac{1}{2\pi }\oint d\phi =n\text{, \ \ }n=1,2,3....,
\end{eqnarray*}%
where we made use of the result of the change of phase in parallel transport
of Schr\"{o}dinger wavefunction in $q$-space,%
\begin{equation*}
d\phi =i\left\langle \alpha ,q\right\vert \frac{\partial }{\partial q}%
\left\vert \alpha ,q\right\rangle \cdot dq\text{,}
\end{equation*}%
where $n\Longrightarrow \alpha $ denotes the discrete energy level\footnote{%
Indeed, if $\left\vert n,q\right\rangle $'s are eigenfunction of momentum
operator, such as for free electron Hamiltonan, then we get exactly the $c$%
-number $p,$ the independent conjugate dynamical variable as proportional to
the connection. B-S quantization makes sense by space compactification such
as a particle in a box. It is this vague strictly momentum viewpoint in the
literature that when naively applied to general cases makes the B-S
condition valid only as a limiting semiclassical approximation in
phase-space, when in fact it has more general validity as demonstrated in
this paper. See also \cite{moll}.}. Furthermore,%
\begin{equation*}
\nabla \phi =\frac{1}{\hbar }\left\langle n,q\right\vert i\hbar \nabla
_{q}\left\vert n,q\right\rangle =\frac{\vec{p}}{\hbar }\equiv :\frac{e}{%
\hbar c}\vec{A},
\end{equation*}%
where we defined $\vec{A}$ as acting like a vector potential. In terms of
the introduced 'vector potential', $\vec{A}$, the B-S quantization condition
now reads%
\begin{eqnarray}
\frac{1}{2\pi \hbar }\oint \vec{p}\cdot d\vec{q} &=&\frac{1}{2\pi }\oint 
\frac{e}{\hbar c}\vec{A}\cdot d\vec{q}  \notag \\
&=&\frac{1}{2\pi }\oint \nabla _{\vec{q}}\phi \left( \vec{q}\right) \cdot d%
\vec{q}=n.  \label{eq1}
\end{eqnarray}

Therefore we arrive at the Bohr-Sommerfeld quantization rules for closed
orbits purely from a geometric considerations, where $\left\langle
n,q\right\vert \nabla _{q_{i}}\left\vert n,q\right\rangle \equiv :\frac{e}{c}%
\vec{A}$ is the Berry connection. It does appear that the B-S quantization
rules have an exact geometric origin, as well as confirming the Planck's
discretization of phase space in units of Planck's constant $h.$ Therefore,
it is expected that B-S condition will play an important role in topological
band theory, as will be shown here. In what follows we give various examples
where the B-S geometric quantization condition enters in the analyses.

If one accounts for vacuum fluctuations due to uncertainty principle, then
the integer $n\Longrightarrow n+\frac{1}{2}$ since $\frac{1}{2}\hbar $ is
the minimal vacuum fluctuation of the action.

\section{The B-S condition and quantization of angular momentum}

An immediate case of a 'closed orbits' deal with either orbital or spin
angular momentum. The Schr\"{o}dinger equation for angular momentum about
the $z$-axis is given by%
\begin{equation*}
i\hbar \frac{\partial }{\partial \phi }\left\vert \psi \left( \phi \right)
\right\rangle =\hat{J}_{z}\left\vert \psi \left( \phi \right) \right\rangle ,
\end{equation*}%
where $\hat{J}_{z}$ is the angular momentum operator for the $z$-component.
Therefore, we obtain the B-S condition as,%
\begin{eqnarray}
\hbar \oint \left\langle \psi \left( \phi \right) \right\vert i\frac{%
\partial }{\partial \phi }\left\vert \psi \left( \phi \right) \right\rangle
d\phi &=&\hbar \oint \nabla \theta \left( \phi \right) d\phi =\hbar 2\pi m, 
\notag \\
\oint \left\langle \psi \left( \phi \right) \right\vert \hat{J}%
_{z}\left\vert \psi \left( \phi \right) \right\rangle d\phi &=&J_{z}\oint
d\phi =2\pi J_{z}  \notag \\
&=&2\pi \hbar m.  \label{quant_Sz}
\end{eqnarray}%
Therefore the eigenvalues of the observable component, $J_{z}=m\hbar \leq
\pm \left\vert \vec{J}\right\vert $, $\left( m=\pm 1,\pm 2,3,...\leq \pm
\left\vert \vec{J}\right\vert \right) $, i.e., quantized in units of Planck
action. This B-S quantization condition has been used by Haldane \cite%
{haldane} in his theory of a new state of matter now known as the Haldane
phase.

\section{The B-S phase factor and Schr\"{o}dinger equation}

The Berry phase and Berry curvature was originally derived by Berry \cite%
{berry} using the Schr\"{o}dinger equation whose Hamiltonian depends on a
time-dependent parameter, $R\left( t\right) $. This specifically made clear
in the case of the Born-Oppenheimer approximation in molecular physics,
where the Berry connection exactly acts like a vector potential in the
effective Schr\"{o}dinger equation for the slow variable \cite{buotbook}.
Therefore, the wavefunction evolves both in time and parameter space.
However, because of the presence of the dynamical phase factor in the
wavefunction, the B-S phase factor is not the only phase factor and hence
the B-S quantization condition, on grounds of wavefunction uniqueness,
cannot be implemented. This is left as integral around a contour in
parameter $R$-space, under time \textit{compactification} (i.e.,
'time-Brillouin zone' or 'temporal box': $t=0$ to $T$) of steady-state
condition. We have%
\begin{equation*}
\left\vert \psi \left( T\right) \right\rangle =\exp \left\{ \frac{-i}{\hbar }%
\int\limits_{0}^{T}dtE_{n}\left( R\left( t\right) \right) \right\} \exp
\left\{ i\gamma _{n}\left( C\right) \right\} \left\vert \psi \left( R\left(
0\right) \right) \right\rangle ,
\end{equation*}%
where the B-S phase or geometrical phase factor is given by, 
\begin{equation*}
\gamma _{n}\left( C\right) =\oint\limits_{C}\ \left\langle n,R\right\vert i%
\frac{\partial }{\partial \vec{R}}\left\vert n,R\right\rangle \cdot d\vec{R}.
\end{equation*}

\section{The B-S condition and Landau-level degeneracy}

Indeed, the quantization condition in terms of the vector potential, Eq. (%
\ref{eq1}) materializes in the method of counting of Planck states in phase
space for the Landau circular orbits. This is the calculation of the
Landau-level (L-L) degeneracy. Classically this may be approximated by%
\begin{equation}
\frac{A}{\pi r^{2}}=\frac{\pi R^{2}}{\pi r^{2}}=\frac{R^{2}}{r^{2}},
\label{eqCLD}
\end{equation}%
where $r$ is the classical radius of Landau orbits in a uniform magnetic
fields and the system area is given by $A=\pi R^{2}$. Equation (\ref{eqCLD})
has the dimensional units of total flux divided by the dimensional units of
'quantum' flux, $\frac{\left( \pi R^{2}\right) B}{\pi \left( \frac{EL}{e}%
\right) }\Longrightarrow \sim \frac{\Phi }{\phi _{o}}$.

Quantum mechanically, the more accurate expression for the degeneracy is $%
\frac{\Phi }{\frac{2\pi \hbar c}{e}}=\frac{\Phi }{\phi _{o}}$. This can be
inferred simply by counting of Planck states ('pixel' of action) in phase
space using Berry's curvature and connection, i.e., magnetic field and
vector potential, respectively. In Gaussian units, we have,%
\begin{eqnarray}
L\text{-}L\text{ }Degeneracy &=&\frac{1}{2\pi \hbar }\iint \vec{\nabla}%
\times \vec{K}\cdot d\vec{a}  \notag \\
&=&\frac{1}{2\pi \hbar }\iint \vec{\nabla}\times \frac{e}{c}\vec{A}\cdot d%
\vec{a}  \notag \\
&=&\frac{1}{2\pi \hbar }\frac{\left\vert e\right\vert }{c}\iint \vec{B}\cdot
d\vec{a}  \notag \\
&=&\frac{\Phi }{\phi _{o}}\ \epsilon \ 
\mathbb{Z}
\notag \\
&=&\frac{1}{2\pi \hbar }\oint \frac{e}{c}\vec{A}\cdot d\vec{q}=n\ \epsilon \ 
\mathbb{Z}
,  \label{b-s_version}
\end{eqnarray}%
where $\Phi $ is the total magnetic flux, and $\phi _{o}$ is the quantum flux%
$=\frac{hc}{\left\vert e\right\vert }$. This is a realization of the B-S
quantization condition given in Eq. (\ref{eq1}). The resulting quantization
of orbital motion leads to edge states and integer quantum Hall effect under
uniform magnetic fields. The general analysis of edge states marks the works
of Laughlin \cite{laughlin}, and Halperin \cite{halperin}.

\section{Peierls phase factor and Wilson loops}

The above argument on confined or bounded motion can be made precise by
noting that localized wavefunctions in energy-band theory with vector
potential, either real electromagnetic or Berry connection, always carry the
so-called Peierls phase factor. This is well-known since the early days of
solid-state physics. Thus, 'bringing' (i.e., using \textit{magnetic
translation operator} defined below) a localized wavefunction around a
closed loop (in modern terminology the so-called Wilson loop or plaquette)
would acquire a total phase factor determined by Eq. (\ref{eq1}),%
\begin{equation*}
\exp \left( \frac{i}{\hbar }\oint\limits_{C}\frac{e}{c}A\cdot dq\right) =1%
\text{ ,}
\end{equation*}%
by virtue of uniqueness. With the contour enclosing a magnetic flux and
dividing the closed trajectory into halves with each half endowed with
opposite chirality, one end up with Aharonov-Bohm effect.

For a Wilson loop, this means that for an electromagnetic vector potential,
we have,%
\begin{eqnarray*}
\frac{i}{\hbar }\oint\limits_{C}\frac{e}{c}A\cdot dq &=&2\pi n\text{, \ }n%
\text{ }\epsilon \ 
\mathbb{Z}
, \\
\frac{i}{2\pi \hbar }\frac{e}{c}\iint \nabla \times A\cdot d\vec{a} &=&n%
\text{, \ }n\text{ }\epsilon \ 
\mathbb{Z}
\\
&=&\frac{\Phi }{\phi _{o}}\ \epsilon \ 
\mathbb{Z}
,
\end{eqnarray*}%
which again follows from the Bohr-Sommerfeld quantization condition.

We note that the Landau orbits are confined in phase space and the amount of
"$h^{3}$-voxels" or more appropriately the number of $h$-pixels of phase
space enclosed by the orbits in units of Planck's constant is quantized,
i.e., $\epsilon \ 
\mathbb{Z}
$. We can even say that any oscillatory and harmonic motion in phase space
entails some interdependence of canonical coordinates so as to enclose an
integral number of Planck's states, i.e., $h$-pixels. In the case of
magnetic fields, this translates to the number of quantum fluxes. The
minimal coupling of canonical momentum under a uniform magnetic field
provides the natural interdependence of the canonical conjugate variables,
where the dependence on the coordinates is only provided by the vector
potential.

\section{B-S quantization of magnetic charge in a Dirac monopole}

The classical Maxwell's equations are

\begin{eqnarray*}
\nabla \cdot E &=&\rho _{E}\text{, \ \ \ \ \ \ \ \ \ \ }\nabla \times E=-%
\frac{\partial B}{\partial t}, \\
\nabla \cdot B &=&0\text{, \ \ \ \ \ \ \ \ \ \ \ \ }\nabla \times B=\frac{%
\partial E}{\partial t}+j_{E}.
\end{eqnarray*}%
If Dirac magnetic monopole exists, giving a magnetic charge density, $\rho
_{m}$, then we have a complete duality of electric and magnetic fields,

\begin{eqnarray}
\nabla \cdot E &=&\rho _{E}\text{, \ \ \ \ \ \ \ \ \ \ }\nabla \times E=-%
\frac{\partial B}{\partial t}-j_{m},  \notag \\
\nabla \cdot B &=&\rho _{m}\text{, \ \ \ \ \ \ \ \ \ \ \ \ }\nabla \times B=%
\frac{\partial E}{\partial t}+j_{E}.  \label{symmetricMBeq}
\end{eqnarray}%
The electric-magnetic duality is characterized by the following replacements,%
\begin{eqnarray*}
E &\Longrightarrow &B\text{, \ \ \ \ }\rho _{E}\Longrightarrow \rho _{m}%
\text{, \ \ \ \ \ }j_{E}\Longrightarrow j_{m}, \\
B &\Longrightarrow &-E\text{, \ \ \ \ }\rho _{m}\Longrightarrow -\rho _{E}%
\text{, \ \ \ \ \ }j_{m}\Longrightarrow -j_{E}.
\end{eqnarray*}%
The four equations in Eq. (\ref{symmetricMBeq}) is condensed into two using
complex fields,%
\begin{eqnarray*}
\nabla \cdot \left( E+iB\right) &=&\left( \rho _{E}+i\rho _{m}\right) , \\
\nabla \times \left( E+iB\right) &=&i\frac{\partial }{\partial t}\left(
E+iB\right) +i\left( j_{E}+ij_{m}\right) ,
\end{eqnarray*}%
which manifest additional symmetry of the electric-magnetic duality.
Although, classical arguments persist against the existence of magnetic
monopole, quantum theory is more favorable of its existence, as will be
shown below.

Here, we can continue to make use of the Peierl's phase-factor arguments to
deduce the existence of Dirac magnetic monopole, instead of the use of
semi-infinitely-long/thin Dirac string originally employed by Dirac \cite%
{dirac}. This done by creating a tetrahedron or bounded $3$-D domain using
the magnetic translation operator to define the surfaces of the bounded
region. The '\textit{pointed tip of the pen}' is served by a localized
function represented by the magnetic Wannier function translated between
lattice sites to define a tetrahedron.

For our purpose, it is convenient to formalize the magnetic translation
operator in the presence of vector potential defined by%
\begin{equation*}
T\left( \vec{q}\right) =\exp \left[ \frac{i}{\hbar }\left( \frac{\hbar }{i}%
\nabla _{\vec{r}}+\frac{e}{c}\vec{A}\left( \vec{r}\right) \right) .\vec{q}%
\right] ,
\end{equation*}%
where $\vec{A}\left( \vec{r}\right) $ is the vector potential, chosen in
symmetric gauge, $\vec{A}\left( \vec{r}\right) =\frac{1}{2}\vec{B}\times 
\vec{r}$, and $\vec{q}$ is the crystal lattice vector.\footnote{%
The continuum limit of all the results can be taken in the end.} $T\left( 
\vec{q}\right) $ generates all the magnetic Wannier functions belonging to a
band index $\lambda $ from a given Wannier function centered at the origin.
Therefore, operating on the lattice-position eigenvector centered at the
origin, we have 
\begin{equation*}
T\left( \vec{q}\right) \left\vert 0,\lambda \right\rangle =\left\vert \vec{q}%
,\lambda \right\rangle .
\end{equation*}%
Consider%
\begin{equation*}
T\left( -\vec{q}_{1}\right) =\exp \left[ -\frac{i}{\hbar }\left( \frac{\hbar 
}{i}\nabla _{\vec{r}}+\frac{e}{c}\vec{A}\left( \vec{r}\right) \right) .\vec{q%
}_{1}\right] .
\end{equation*}%
Using the Baker--Campbell-Hausdorff (BCH) formula, we have%
\begin{eqnarray*}
&&\exp \left[ -\frac{i}{\hbar }\left( \frac{\hbar }{i}\nabla _{\vec{r}}+%
\frac{e}{c}\vec{A}\left( \vec{r}\right) \right) .\vec{q}_{1}\right] \\
&=&\exp \left[ -\frac{i}{\hbar }\left( \frac{e}{c}\vec{A}\left( \vec{r}%
\right) \right) .\vec{q}_{1}\right] \exp \left[ -\frac{i}{\hbar }\left( 
\frac{\hbar }{i}\nabla _{\vec{r}}\right) .\vec{q}_{1}\right] .
\end{eqnarray*}%
Using the localized wavefunction, $w_{\lambda }\left( r,0\right) $, this
means that%
\begin{equation}
T\left( -\vec{q}_{1}\right) w_{\lambda }\left( r,0\right) =\exp \left( -%
\frac{i}{\hbar }\frac{e}{c}\vec{A}\left( \vec{r}\right) .\vec{q}_{1}\right)
w_{\lambda }\left( \vec{r}-\vec{q}_{1}\right) ,  \label{peierls_phase}
\end{equation}%
where now the displaced localized wavefunction is now \textit{centered} at
the lattice point $\vec{q}$. We define $\tilde{w}_{\lambda }\left( \vec{r}-%
\vec{q}\right) $ as the magnetic Wannier function centered at $\vec{q}$.
Because we are translating a localized Wannier function, we need to perform
negative lattice vector shift to obtain a Wannier function centered in a new
resultant vector lattice point. The exponential factor of Eq. (\ref%
{peierls_phase}) is the so-called \textit{Peierls phase factor}.

The $T\left( \vec{q}\right) $-algebra is determined from BCH formula,%
\begin{eqnarray*}
T\left( -\vec{q}_{1}\right) T\left( -\vec{q}_{2}\right)  &=&\exp \left[ -%
\frac{i}{\hbar }\left( \frac{\hbar }{i}\nabla _{\vec{r}}+\frac{e}{c}\vec{A}%
\left( \vec{r}\right) \right) .\vec{q}_{1}\right] \exp \left[ -\frac{i}{%
\hbar }\left( \frac{\hbar }{i}\nabla _{\vec{r}}+\frac{e}{c}\vec{A}\left( 
\vec{r}\right) \right) .\vec{q}_{2}\right]  \\
&=&\exp \left( -\frac{ie}{\hbar c}\vec{A}\left( \vec{q}_{1}\right) .\vec{q}%
_{2}\right) T\left( -\left( \vec{q}_{1}+\vec{q}_{2}\right) \right)  \\
&=&\exp \left( -\frac{ie}{\hbar c}\vec{B}.\frac{1}{2}\left( \vec{q}%
_{1}\times \vec{q}_{2}\right) \right) T\left( -\left( \vec{q}_{1}+\vec{q}%
_{2}\right) \right) .
\end{eqnarray*}%
Now consider additional magnetic translation by $\vec{q}_{3}$ We obtain%
\begin{eqnarray}
&&\left[ \left( T\left( -\vec{q}_{1}\right) T\left( -\vec{q}_{2}\right)
\right) \right] \ T\left( -\vec{q}_{3}\right)   \notag \\
&=&\exp \left( -\frac{ie}{\hbar c}\vec{B}.\frac{1}{2}\left( \vec{q}%
_{1}\times \vec{q}_{2}\right) \right) \exp \left( -\frac{ie}{\hbar c}\vec{B}.%
\frac{1}{2}\left( \vec{q}_{1}+\vec{q}_{2}\right) \times \vec{q}_{3}\right)
T\left( -\left( \vec{q}_{1}+\vec{q}_{2}+\vec{q}_{3}\right) \right)   \notag
\\
&=&\exp \left( -\frac{ie}{\hbar c}\vec{B}.\left[ \frac{1}{2}\left( \vec{q}%
_{1}\times \vec{q}_{2}\right) +\frac{1}{2}\left( \vec{q}_{1}+\vec{q}%
_{2}\right) \times \vec{q}_{3}\right] \right) T\left( -\left( \vec{q}_{1}+%
\vec{q}_{2}+\vec{q}_{3}\right) \right) .  \label{first2}
\end{eqnarray}%
On the other hand, using a different bilinear grouping, we can have

\begin{equation}
T\left( -\vec{q}_{1}\right) \ \left[ T\left( -\vec{q}_{2}\right) T\left( -%
\vec{q}_{3}\right) \right] =\exp \left( -\frac{ie}{\hbar c}\phi _{23}\right)
\exp \left( -\frac{ie}{\hbar c}\phi _{1\left( 2+3\right) }\right) T\left(
-\left( \vec{q}_{1}+\vec{q}_{2}+\vec{q}_{3}\right) \right) .  \label{last2}
\end{equation}%
By virtue of the associative properties of the quantum operators $T\left( 
\vec{q}\right) $, The RHS of Eq. (\ref{first2}) must be equal to the RHS of
Eq. (\ref{last2}). We obtain four surfaces enclosing a tetrahedron defined
by the phase,

\begin{eqnarray}
&&T\left( -\left( \vec{q}_{1}+\vec{q}_{2}+\vec{q}_{3}\right) \right)  \notag
\\
&=&\exp \left( -\frac{ie}{\hbar c}\left[ \phi _{12}+\phi _{\left( 1+2\right)
3}-\phi _{23}+\phi _{1\left( 2+3\right) }\right] \right) T\left( -\left( 
\vec{q}_{1}+\vec{q}_{2}+\vec{q}_{3}\right) \right) ,  \label{Trans_teta}
\end{eqnarray}%
where $\left[ \phi _{12}+\phi _{\left( 1+2\right) 3}-\phi _{23}+\phi
_{1\left( 2+3\right) }\right] =\Phi $ is the total magnetic flux emerging
out of bounded domain enclosed by the surfaces formed by tetrahedron.
Writing explicitly the total flux, $\Phi $, emerging out of tetrahedron, 
\begin{eqnarray}
\Phi &=&\vec{B}\cdot \frac{1}{2}\left[ \vec{q}_{2}\times \vec{q}_{1}+\left( 
\vec{q}_{3}\times \left( \vec{q}_{1}+\vec{q}_{2}\right) \right) +\left( \vec{%
q}_{2}\times \vec{q}_{3}\right) +\left( \vec{q}_{1}\times \left( \vec{q}_{2}+%
\vec{q}_{3}\right) \right) \right]  \notag \\
&=&\vec{B}\cdot \left[ \frac{1}{2}%
\begin{array}{c}
\left( \vec{q}_{1}+\vec{q}_{2}\right) \times \vec{q}_{1}+\left( \left( \vec{q%
}_{1}+\vec{q}_{2}+\vec{q}_{3}\right) \times \left( \vec{q}_{1}+\vec{q}%
_{2}\right) \right) \\ 
+\left( \vec{q}_{2}\times \left( \vec{q}_{2}+\vec{q}_{3}\right) \right)
+\left( \vec{q}_{1}\times \left( \vec{q}_{1}+\vec{q}_{2}+\vec{q}_{3}\right)
\right)%
\end{array}%
\right] .  \label{flux_in_out}
\end{eqnarray}%
This means that the phase of Eq. (\ref{Trans_teta}) is zero, as expected
since application of the resulting total magnetic translation, $T\left( -%
\vec{q}_{1}\right) T\left( -\vec{q}_{2}\right) T\left( -\vec{q}_{3}\right) $
applied to a localized Wannier function, $w\left( \vec{r},o\right) $, should
be equivalent to 
\begin{eqnarray}
&&T\left( -\left( \vec{q}_{1}+\vec{q}_{2}+\vec{q}_{3}\right) \right) w\left( 
\vec{r},o\right)  \notag \\
&=&\exp \left( -\frac{i}{\hbar }\frac{e}{c}A\left( \vec{r}\right) .\left( 
\vec{q}_{1}+\vec{q}_{2}+\vec{q}_{3}\right) \right) w\left( \vec{r}-\left( 
\vec{q}_{1}+\vec{q}_{2}+\vec{q}_{3}\right) \right)  \notag \\
&=&\exp \left( -\frac{i}{\hbar }\frac{e}{c}A\left( \vec{r}\right) .\vec{q}%
\right) w\left( \vec{r}-\vec{q}\right) ,  \label{peierls_phase2}
\end{eqnarray}%
where $\vec{q}\equiv \left( \vec{q}_{1}+\vec{q}_{2}+\vec{q}_{3}\right) $ in
the last line consistent with Eq. (\ref{peierls_phase}).

\begin{figure}[h]
\centering
\includegraphics[width=1.8226in]{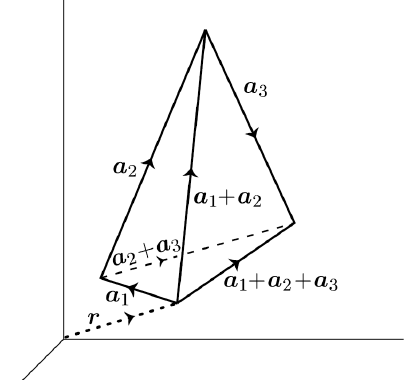}
\caption{The tetrahedron through which the emerging magnetic fluxes are
determined, demanding a source of magnetic flux within the enclosed domain
if the total flux do not add to zero. [Figure reproduced from Ref. 
\protect\cite{jackiw}]. Here $a_{i}\Longrightarrow \vec{q}_{i}$ in the text. 
}
\label{fig1}
\end{figure}

The total phase of Eq. (\ref{Trans_teta}) suggest the following integrals%
\begin{eqnarray*}
\frac{ie}{\hbar c}\Phi &\Longrightarrow &\frac{ie}{\hbar c}\oint\limits_{S}%
\vec{B}\cdot d\vec{s} \\
&=&\frac{ie}{\hbar c}\iiint\limits_{V_{\text{tetrahedron}}}dv\ \vec{\nabla}%
\cdot \vec{B},
\end{eqnarray*}%
where $d\vec{s}$ denotes the incremental surface in the $r$-coordinates (not
the lattice coordinates). Since $\vec{\nabla}\cdot \vec{B}=0$ then $\Phi =0$%
. However, there is another solution for the phase, which can be written in
detail as%
\begin{eqnarray}
&&\frac{ie}{\hbar c}\Phi  \notag \\
&=&\frac{ie}{\hbar c}\vec{B}\cdot \left[ 
\begin{array}{c}
\left( \vec{q}_{1}+\vec{q}_{2}\right) \times \vec{q}_{1}+\left( \left( \vec{q%
}_{1}+\vec{q}_{2}+\vec{q}_{3}\right) \times \left( \vec{q}_{1}+\vec{q}%
_{2}\right) \right) \\ 
+\left( \vec{q}_{2}\times \left( \vec{q}_{2}+\vec{q}_{3}\right) \right)
+\left( \vec{q}_{1}\times \left( \vec{q}_{1}+\vec{q}_{2}+\vec{q}_{3}\right)
\right)%
\end{array}%
\right]  \notag \\
&=&i2\pi n,  \label{nonzero}
\end{eqnarray}%
which clearly exhibit an expression with all the magnetic fluxes emerging
from the tetrahedron. Without affecting the validity of Eq. (\ref%
{peierls_phase2}), the total phase of Eq. (\ref{Trans_teta}) for all
emerging magnetic flux of Eq. (\ref{nonzero}) must satisfy the following
expression%
\begin{eqnarray*}
&&\frac{ie}{\hbar c}\Phi \\
&=&\frac{ie}{\hbar c}\oint\limits_{S}\vec{B}\cdot d\vec{s}=\frac{ie}{\hbar c}%
\iiint\limits_{V_{\text{tetrahedron}}}dv\ \vec{\nabla}\cdot \vec{B} \\
&=&i2\pi n\text{, \ }n\ \epsilon \ 
\mathbb{Z}
\text{, \ \ \ }n=0,\pm 1,\pm 1,.....,
\end{eqnarray*}%
if there is a magnetic charge $e_{m}$, '$+$' for monopoles and '$-$' for
anti-monopoles. The magnetic field $\vec{B}$ is such that 
\begin{eqnarray*}
\vec{B} &=&e_{m}\frac{\vec{r}}{r^{3}} \\
\vec{\nabla}\cdot \vec{B} &=&4\pi e_{m}\delta ^{3}\left( \vec{r}\right)
\Longrightarrow \iiint\limits_{V_{\text{tetrahedron}}}dv\ \vec{\nabla}\cdot 
\vec{B}=4\pi e_{m} \\
\nabla ^{2}\phi _{m} &=&4\pi e_{m}\delta ^{3}\left( \vec{r}\right)
\end{eqnarray*}%
then it follows that the magnetic charge, $e_{m}$, is quantized,%
\begin{equation*}
\frac{ie}{\hbar c}\iiint\limits_{V_{\text{tetrahedron}}}dv\ \vec{\nabla}%
\cdot \vec{B}=\frac{ie}{\hbar c}4\pi e_{m}=i2\pi n
\end{equation*}%
which yields the Dirac magnetic monopole,%
\begin{equation*}
e_{m}=n\left( \frac{\hbar c}{2e}\right) =\frac{1}{4\pi }n\phi _{o},
\end{equation*}%
which was originally given by Dirac by his thought experiment of
semi-infinite thin magnetic string, where $\left( \frac{\hbar c}{2e}\right) =%
\frac{e}{2\alpha }=\frac{137}{2}e$ ($\alpha $ is the fine structure constant 
$=\frac{e^{2}}{\hbar c}$) is the unit magnetic charge. Now therefore, 
\begin{equation*}
\iiint\limits_{V_{\text{tetrahedron}}}dv\ \vec{\nabla}\cdot \vec{B}=0\text{,
and }\iiint\limits_{V_{\text{tetrahedron}}}dv\text{ }\vec{\nabla}\cdot \vec{B%
}\neq 0,
\end{equation*}%
would be physically consistent if $\vec{\nabla}\cdot \vec{B}=0$ outside a
singularity and $\vec{\nabla}\cdot \vec{B}=e_{m}$ only at the core. This
allows us to propose that%
\begin{eqnarray*}
\vec{B} &=&e_{m}\frac{\vec{r}}{r^{3}}, \\
\vec{\nabla}\cdot \vec{B} &=&4\pi e_{m}\delta ^{3}\left( \vec{r}\right) , \\
\nabla ^{2}\phi _{m} &=&4\pi e_{m}\delta ^{3}\left( \vec{r}\right) ,
\end{eqnarray*}%
where $\phi _{m}$ is the magnetic monopole potential. $\vec{\nabla}\cdot 
\vec{B}=\nabla ^{2}\phi _{m}=0$ if the singular point $\vec{r}=0$ is
excluded. We have%
\begin{equation*}
\iiint\limits_{\delta V_{core}}dv\text{ }\vec{\nabla}\cdot \vec{B}=n\left( 
\frac{e}{2\alpha }\right) ,
\end{equation*}%
where $\delta V_{core}$ is an infinitisimal domain containing the
singularity at $\vec{r}=0$ or the quantum flux is emerging through a
spherical surface surrounding the point of singularity. Clearly a magnetic
monopole is \textit{infinitely localized} and therefore must be hugely
massive. Thus, it requires very high-energy not currently readily available
in experimental physics to detect it. Moreover, it may always occur in
pairs, namely, monopole and anti-monopole pair (i.e., north and south pole,
similar to vortex anti-vortex pairs of the $X$-$Y$ model of spin systems) of
infinitesimal size which conspire to evade experimental detection as
magnetic dipoles.

\section{The X-Y model of spin systems and B-S condition}

The $X$-$Y$ model is a sort of a generalization of the Ising model, i.e.,
instead of discrete $\pm 1$ spin value one place a spin rotor at each site
which can point in any direction in a two-dimensional plane. There is a
theorem based on Landau symmetry breaking arguments that such $2$-D systems
are not expected to exhibit long-range order due to transverse fluctuations 
\cite{mermin}. However, this $2$-D system possess quasi-long-range order in
finite-size systems at very low temperatures. This is the so-called
Berezinskii-Kosterlitz-Thouless (BKT) transition \cite{thouless, bkt},
marked by the occurrence of bound vortex-antivortex pairs the at low
temperatures to unbound or isolated vortices and antivortices above some
critical temperatures.

In 1972, Kosterlitz and Thouless (KT) made a complete identification of a
new type of phase transition in $2$-D systems where topological defects in
the form of vortices and antivortices play a crucial role. For their work
they were awarded the Nobel Prize in 2016. Here we are mainly concern on how
the B-S condition governs the nature of vortices of the $X$-$Y$ model of
spin systems. After the work of KT strong interest towards phase transition
without symmetry breaking become mainstream. It was realized in passing that
long before, the regular liquid-gas transition does not break symmetry, this
was recognized without given much significance. Later, Polyakov extended the
work of KT to gauge theories, for example, $2$+$1$ "compact" QED has a
gapped spectrum in the IR due to topological excitations \cite{polyakov}.
The SU(N) Thirring model has a fermions condensing with finite mass in the
IR without breaking the chiral symmetry of the theory \cite{witten}. It was
also further extended by in the context of $2$-D melting of crystalline
solids \cite{nelson} leading to a new liquid crystalline hexatic phase.
Another violation of Landau symmetry breaking arguments is exemplified by
the Haldane phase transition \cite{haldane}.

\subsection{\qquad The Hamiltonian for the $X$-$Y$ model of spin system}

The $X$-$Y$ model is a generalization of the Ising model, where Ising spins $%
\sigma =\pm 1$ are replaced by planar vector rotors of unit length, which
can point in an arbitrary direction within the $X$-$Y$ plane,%
\begin{equation*}
\vec{n}=\left( 
\begin{array}{c}
\cos \varphi \\ 
\sin \varphi%
\end{array}%
\right) \Longrightarrow \left\vert \vec{n}\right\vert =1.
\end{equation*}%
The Hamiltonian is given by%
\begin{eqnarray}
H &=&-J\sum\limits_{\left\langle i,j\right\rangle }\vec{n}_{i}\cdot \vec{n}%
_{j}  \notag \\
&=&-J\sum\limits_{\left\langle i,j\right\rangle }\cos \left( \vartheta
_{i}\right) \cos \left( \vartheta _{j}\right) +\sin \left( \vartheta
_{i}\right) \sin \left( \vartheta _{j}\right)  \notag \\
&=&-J\sum\limits_{\left\langle i,j\right\rangle }\cos \left( \vartheta
_{i}-\vartheta _{j}\right) .  \label{coupl_sites}
\end{eqnarray}%
The cosine function can be expanded in powers of $\left( \vartheta
_{i}-\vartheta _{j}\right) $,%
\begin{equation}
H=-J\sum\limits_{\left\langle i,j\right\rangle }\left[ 1-\frac{1}{2}\left(
\vartheta _{i}-\vartheta _{j}\right) ^{2}+O\left[ \left( \vartheta
_{i}-\vartheta _{j}\right) ^{4}\right] ...\right] .  \label{expand}
\end{equation}%
In the continuum limit, Eq. (\ref{expand}) leads to the following
Hamiltonian studied by BKT,%
\begin{equation}
H\simeq -2JN+\frac{J}{2}\int d^{2}r\left\vert \nabla \theta \left( r\right)
\right\vert ^{2},  \label{contH}
\end{equation}%
where $-2JN\equiv :E_{o}$ is the energy of the system when all $N$ spin
rotors are aligned, and $\theta \left( r\right) $ labels the angle of the
rotors at each point in the $X$-$Y$ plane. The partition function in this
continuum limit must account for all possible configurational function $%
\left\{ \theta \left( r\right) \right\} $. We are thus lead to a \textit{%
functional integral} of the partition function, $Z$,%
\begin{equation}
Z\Longrightarrow \int D\left[ \theta \left( r\right) \right] \exp \left[
-\beta \left( E_{o}+\frac{J}{2}\int d^{2}r\left\vert \nabla \theta \left(
r\right) \right\vert ^{2}\right) \right] .  \label{func_integ}
\end{equation}

\begin{remark}
The $X$-$Y$ model of spin system also serves as a beautiful model of Berry
connection and Berry curvature in a more explicit and much simpler form.
This naturally leads us to the B-S quantization condition for the vortex
solutions given by KBT. It is therefore expected that this model will have
an impact not only in statistical physics but also in quantum field theory
and subject to generalizations. For example, if we write the unit spin
vector by employing the Dirac ket and bra notations, we have%
\begin{equation*}
\left\vert \vec{n}\right\rangle =e^{-i\theta \left( r\right) }
\end{equation*}%
then%
\begin{equation*}
\left\langle \vec{n}\right\vert i\frac{\partial }{\partial r}\left\vert \vec{%
n}\right\rangle =\nabla \theta \left( r\right)
\end{equation*}%
which is the Berry connection. This site connection is indeed better
appreciated by the way the sites are coupled in Eq. (\ref{coupl_sites}).
This remark serves as advanced view, as it relates to the B-S quantization,
of the discussions that follow.
\end{remark}

\subsubsection{ Vortices as solutions}

To simplify the calculation of the functional integral of Eq. (\ref%
{func_integ}), we use perturbative techniques which allows us to make use of
the saddle point approximation. This entails expansion of the functional in
terms of small fluctuations, $\delta \theta $, around the minimum of $H$ at $%
\theta _{o}$. Therefore, we need%
\begin{equation}
\left. \frac{\delta H\left[ \theta \right] }{\delta \theta \left( r\right) }%
\right\vert _{\theta =\theta _{o}}=0.  \label{min}
\end{equation}%
Then the field configurations are approximated, $\theta \left( r\right)
\simeq \theta _{o}+\delta \theta \left( r\right) $. We have for the
partition finction,%
\begin{equation*}
Z\simeq e^{-\beta E_{o}}\sum\limits_{\theta _{o}}\int D\left[ \delta \theta %
\right] \exp \left\{ -\beta \left( H\left[ \theta _{o}\right] +\frac{1}{2}%
\iint dr_{1}dr_{2}\frac{\delta ^{2}H\left[ \theta \right] }{\delta \theta
\left( r_{1}\right) \delta \theta \left( r_{2}\right) }\delta \theta \left(
r_{1}\right) \delta \theta \left( r_{2}\right) \right) \right\} .
\end{equation*}%
Here, it is assumed that $H\left[ \theta _{o}\right] $ dominates and the
rest are fluctuations. With $H\left[ \theta \right] $ given by%
\begin{equation*}
H\left[ \theta \right] \equiv \frac{J}{2}\int d^{2}r\left\vert \nabla \theta
\left( r\right) \right\vert ^{2}.
\end{equation*}%
The minimum is determined from the solutions of Eq. (\ref{min}), which
correspond to solving the following equation,%
\begin{equation}
\nabla ^{2}\theta \left( r\right) =0.  \label{laplace}
\end{equation}%
Although the analysis of the solutions to Eq. (\ref{laplace}) proceeds
classically, it has accurate analogy to quantum physics and contains the
elements of Berry phase, Berry curvature and B-S quantization condition.
Indeed, the $X$-$Y$ model has some resemblance to the potential flow in
two-dimensional hydrodynamics \footnote{%
The $X$-$Y$ model has some resemblance to the potential flow in two
dimensional hydrodynamics. There, the Laplace equation, $\nabla ^{2}\varphi
\left( x\right) =0$, is thoroughly analyzed for vortex solutions and the
boundary condition given is the so-called \textit{circulation integral}.}.
There, the Laplace equation, $\nabla ^{2}\varphi \left( x\right) =0$, is
thoroughly analyzed for vortex solutions. Moreover, the $X$-$Y$ model is
reminiscent of the harmonic oscillator which also serves as a perfect
classical forebear of the annihilation and creation operator, or ladder
operator formalism, in quantum mechanics.

\begin{figure}[h]
\centering
\includegraphics[width=2.21175in]{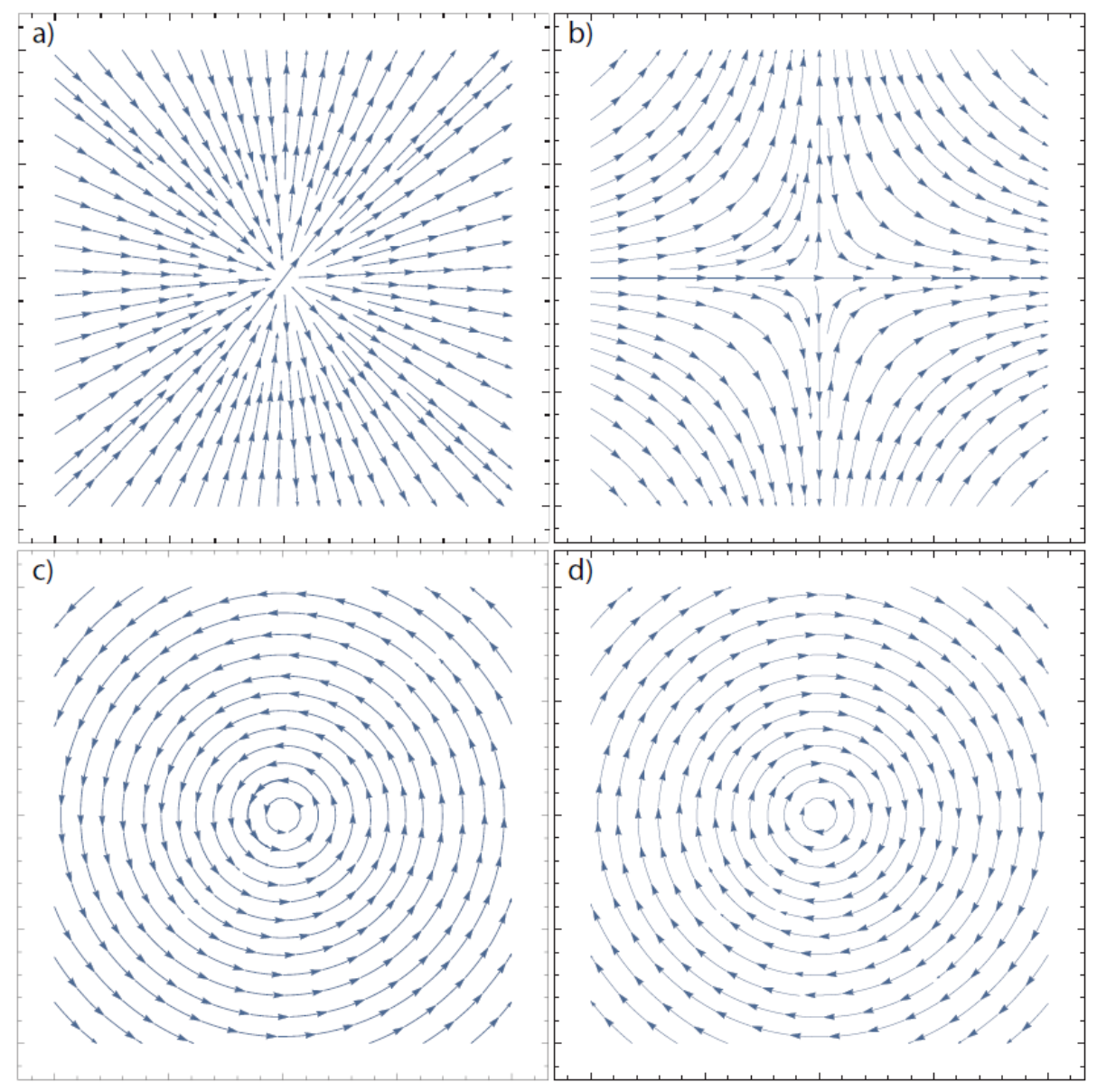}
\caption{ a) Streamlines for the vortex $\protect\theta _{+}$. b)
Streamlines for the antivortex $\protect\theta _{-}$. c) The vortex's
gradient, $\protect\nabla \protect\theta _{+}$. d) The antvortex's gradient, 
$\protect\nabla \protect\theta _{-}$. The contour integral of $\protect%
\nabla \protect\theta _{+}$ is quantized and stands for B-S quantization
condition. [Reproduced from Ref. \protect\cite{ahmed}]}
\label{gradient}
\end{figure}

\begin{figure}[h]
\centering
\includegraphics[width=1.0339in]{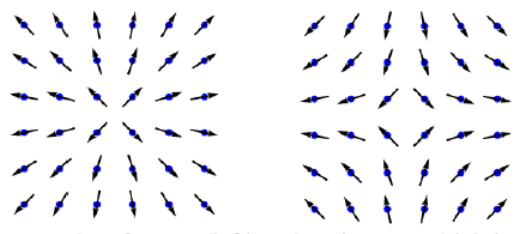}
\caption{Spin configurations for vortex (right) and anti-vortex (left) 
\protect\cite{royal}. }
\label{vortantiv}
\end{figure}

\textbf{T}he two types of solutions to $\nabla ^{2}\theta \left( r\right) =0$%
,

\begin{eqnarray}
(i)\text{ }\theta \left( r\right) &=&\text{constant}\Longrightarrow \text{%
ground state}\Longrightarrow E_{o},  \notag \\
(ii)\text{ }\theta \left( r\right) &=&\text{vortex solutions.}  \label{xybc}
\end{eqnarray}%
The vortex solution is given by%
\begin{equation*}
\theta _{\pm }=\pm \arctan \left( \frac{y-b}{x-a}\right) .
\end{equation*}%
Let%
\begin{equation*}
\theta _{\pm }=\pm \arctan w,
\end{equation*}%
\begin{equation*}
\theta _{\pm }^{\prime }=\pm \frac{w^{\prime }}{1+w^{2}}.
\end{equation*}%
Then we have,%
\begin{equation*}
1+w^{2}=1+\left( \frac{y-b}{x-a}\right) ^{2}=\frac{1}{\left( x-a\right) ^{2}}%
\left( \left( x-a\right) ^{2}+\left( y-b\right) ^{2}\right) ,
\end{equation*}%
\begin{equation*}
\frac{\partial w}{\partial x}=\frac{\partial }{\partial x}\left( \frac{y-b}{%
x-a}\right) =\left( y-b\right) \frac{\partial }{\partial x}\left( \frac{1}{%
x-a}\right) =-\frac{\left( y-b\right) }{\left( x-a\right) ^{2}},
\end{equation*}%
\begin{equation*}
\frac{\partial w}{\partial y}=\frac{\partial }{\partial y}\left( \frac{y-b}{%
x-a}\right) =\frac{1}{x-a}\frac{\partial }{\partial y}\left( \left(
y-b\right) \right) =\frac{1}{\left( x-a\right) }.
\end{equation*}%
The circulation velocities are given by $v=\left( u,v\right) $%
\begin{eqnarray*}
u &=&\frac{-\frac{\left( y-b\right) }{\left( x-a\right) ^{2}}}{\frac{1}{%
\left( x-a\right) ^{2}}\left( \left( x-a\right) ^{2}+\left( y-b\right)
^{2}\right) }=-\frac{\left( y-b\right) }{\left( \left( x-a\right)
^{2}+\left( y-b\right) ^{2}\right) }, \\
v &=&\frac{\frac{1}{\left( x-a\right) }}{\frac{1}{\left( x-a\right) ^{2}}%
\left( \left( x-a\right) ^{2}+\left( y-b\right) ^{2}\right) }=\frac{x-a}{%
\left( \left( x-a\right) ^{2}+\left( y-b\right) ^{2}\right) }.
\end{eqnarray*}%
These results are depicted in Fig. (\ref{gradient})-(\ref{vortantiv}).

The vortex \ solution $\theta _{+}$ corresponds to a vortex charge $+1$, and 
$\theta _{-}$ corresponds to antivortex charge $-1$. Both $\theta _{+}$ and $%
\theta _{-}$ are singular at $\left( x,y\right) =\left( a,b\right) $.

\begin{figure}[h]
\centering
\includegraphics[width=1.8226in]{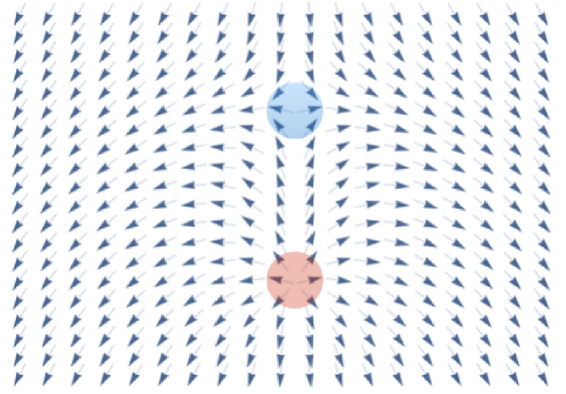}
\caption{Spin configuration for bound vortex-antivortex pair. Color red for
vortex core and color blue for antivortex core \protect\cite{royal}.}
\label{bound}
\end{figure}

The bound vortex-antivortex pair is depicted in Fig. (\ref{bound}).

\subsubsection{B-S quantization of $X$-$Y$ model vortex charges}

For the $X$-$Y$ model of spin system, the quantization of vortex charge, in
contrast to the magnetic monopole charge, usually proceeds classically in a
form of a simple \textit{boundary condition}, the so-called \textit{%
circulation integral (borrowed from }$\mathit{2}$\textit{-D hydrodynamics)}
belonging to the domain of integers, $%
\mathbb{Z}
$. Since the existence of a vortex is based on the existence of a
singularity, the vortex charge can be singly charged, $\pm 1$ or multiply
charged, $\pm m$. The boundary condition for vortex solution is given by Eq.
(\ref{xybc}). The condition, $\oint\limits_{C}\vec{\nabla}\theta \cdot d\vec{%
l}=2\pi m$ basically follows from the single valuedness of $\vec{n}=\left( 
\begin{array}{c}
\cos \theta \\ 
\sin \theta%
\end{array}%
\right) $. The number $m\in Z$ is also called the winding number or vortex
charge (by identifying with the potential problem in electrostatics, 
\begin{equation*}
\int\limits_{V}\vec{\nabla}\cdot \vec{\nabla}\theta =\int_{V}\vec{\nabla}%
\cdot \vec{E}=\int \vec{E}\cdot \vec{n}\ dS,
\end{equation*}%
which is equal to the electric charge enclosed by the surface).
Mathematically speaking, here our order parameter manifold is $S^{1}$ $%
\Longrightarrow \pi _{1}\left( S^{1}\right) =Z$.

Only when there is a singularity would there be a point charge, or vortex
core in the present instance given by%
\begin{equation}
\frac{1}{2\pi }\oint\limits_{\partial S}\vec{\nabla}\theta \cdot d\vec{l}=%
\frac{1}{2\pi }\int\limits_{S}\left( \vec{\nabla}\times \vec{\nabla}\theta
\right) \cdot dS=m,  \label{B-Sanalog}
\end{equation}%
which is our B-S quantization analogy for the $X$-$Y$ model.

\subsubsection{Energetics of \ $X$-$Y$ model}

The energy cost of a single vortex has been shown to be, 
\begin{equation*}
U_{v}=\pi J\ln \frac{L}{a},
\end{equation*}%
where $L$ is the size of the system and $a$ is the lattice constant.
Whereas, the entropy of a vortex is given by%
\begin{equation*}
S_{v}=2k_{B}\ln \frac{L}{a}.
\end{equation*}%
Thus the free energy cost due to presence of vortex, which give the
competition between order and disorder is%
\begin{eqnarray*}
\Delta F &=&U_{v}-TS_{v} \\
&=&\left( \pi J-2k_{B}T\right) \ln \frac{L}{a}
\end{eqnarray*}%
The BKT transition occurs when $\Delta F\Longrightarrow 0$. This give the
transition temperature, $T_{KBT}=\frac{\pi J}{2k_{B}}$. For $T<T_{KBT}$, the
system is unstable against the formation of vortex-antivotex pair. The
energy of the vortex-antivortex pair is%
\begin{equation*}
E_{pair}=\pi J\ln \frac{r}{a},
\end{equation*}%
where $r$ is the separation of the pair. This is much less than the energy
of a single vortex in the limit of large system size. For $T>T_{KBT}$, the
pairs become unbounded.

\subsection{B-S quantization condition in Haldane phase}

It is worth mentioning that quantization of the internal precession
frequency of spin-$1$ chain soliton studied by Haldane \cite{haldane}, i.
e., soliton angular momentum, $S_{Z}$, follows that of Eq. (\ref{quant_Sz}).
The gapped ground state of odd integer-spin chain is now called the Haldane
phase, a new topological state of matter \cite{wen}. The Haldane phase, also
called a symmetry protected topological state, is viewed as a short-range
entangled which bears some similarity to the Kosterlitz-Thouless vortex
solution of the $X$-$Y$ model of spin. We will not go into the details of
Haldane phase as this will take us far from the scope of this paper.

\subsubsection{Haldane model}

A Chern insulator in a honeycomb lattice that exhibits quantum Hall effect 
\cite{armas} in the absence of external magnetic fields was proposed by
Haldane \cite{haldane2} as a basic theoretical representation for the
quantum anomalous Hall effect. The calculation of the Berry connection
relies on the solution of the eigenfunctions and eigenvalues of the
tight-binding Hamiltonian of spinless honeycomb lattice. We refer the
interested readers to some of the references listed in this paper.

\section{B-S quantization condition in superfluids}

For superfluid the B-S quantization condition is not implemented as a
boundary condition but as a real B-S quantization condition. Superfluid
maybe viewed as a classical complex-valued matter field with emergent
constant of motion, the topological order. Like the $X$-$Y$ model, the
complex matter field of Bose-Einstein condensate, $\psi \left( r\right) $,
is given by%
\begin{equation*}
\psi \left( r\right) =\left\vert \psi \left( r\right) \right\vert e^{i\Phi
\left( r\right) }.
\end{equation*}%
Likewise the superfield velocity field is given by%
\begin{equation*}
v_{s}=\gamma \nabla \Phi ,
\end{equation*}%
where $\gamma =\frac{\hbar }{m}$ is some constant. Here, the circulation is
effectively quantized using the B-S quantization condition, i.e.,%
\begin{eqnarray*}
\frac{m}{2\pi \hbar }\oint\limits_{C}\vec{v}_{s}\cdot d\vec{l} &=&\frac{%
\gamma m}{2\pi \hbar }\oint\limits_{C}\nabla \Phi \cdot d\vec{l} \\
&=&2\pi \frac{\hbar }{m}\frac{m}{2\pi \hbar }\times \text{integer} \\
&=&\func{integer}\epsilon \ 
\mathbb{Z}
.
\end{eqnarray*}

\begin{figure}[h!]
\centering
\includegraphics[width=0.94785in]{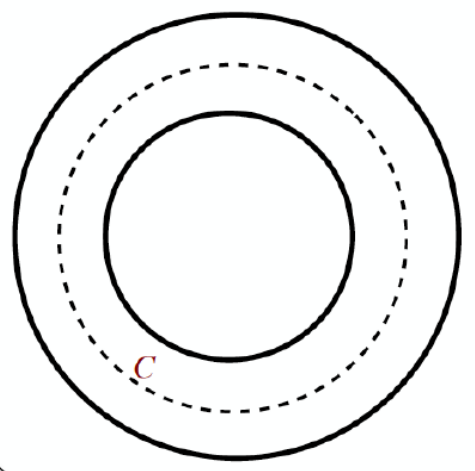}
\caption{An integral contour within the annulus of a superfluid container.
[Reproduced from Ref. \protect\cite{deng}].}
\label{annuluscontour}
\end{figure}

\begin{figure}[h]
\centering
\includegraphics[width=2.2572in]{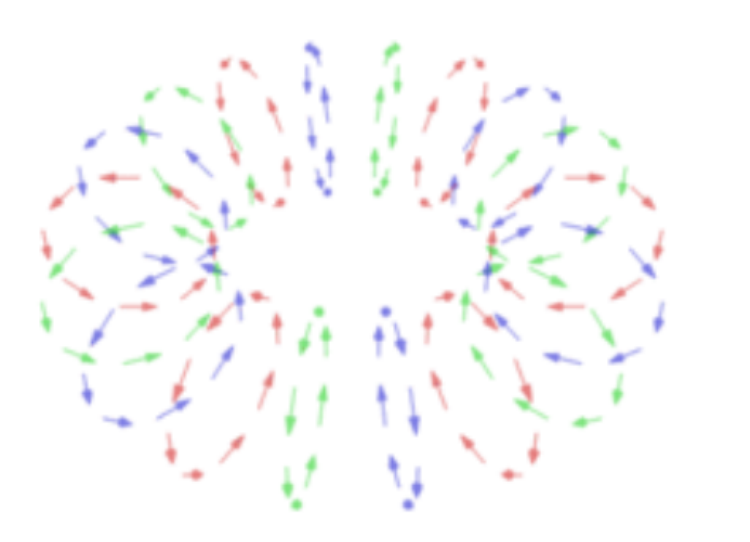}
\caption{A closed quantized vortex ring. [Reproduced from Ref. \protect\cite%
{deng}]}
\label{figrings}
\end{figure}

\begin{figure}[h]
\centering
\includegraphics[width=2.9611in]{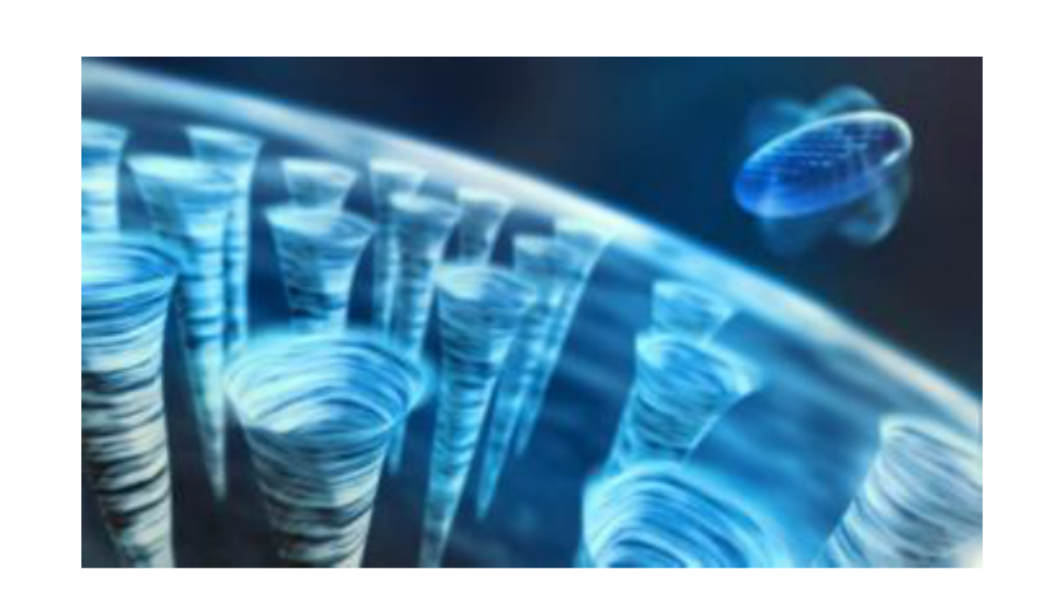}
\caption{An open vortex chains ending at the surface. [Reproduced from Ref. 
\protect\cite{deng}]}
\label{figopenchains}
\end{figure}

Whereas the $X$-$Y$ model is confined to $2$-D, superfluid in $3$-D behave
more distinctly: In $3$-D the superfluid quantized vortices form a
metastable closed ring or open chain ending at the surface, shown in Figs. (%
\ref{annuluscontour})-(\ref{figopenchains}), respectively.

\subsection{B-S quantization of vortices in superconductors}

A well-known property of superconductors is that they expel magnetic fields,
the so-called Meissner effect. In some cases, for sufficiently strong
magnetic fields however, it will be energetically favorable for the
superconductor to form a lattice of quantum vortices through the
superconductor each which carry quantized magnetic flux. A superconductor
that can support vortex lattices is called a type-II superconductor;
vortex-quantization in superconductors is general.

The B-S quantization of vortices is a direct $U\left( 1\right) $ theory of
connection, gauge or vector potential. We have, 
\begin{eqnarray*}
\frac{1}{2\pi \hbar }\frac{e}{c}\iint\limits_{S}\vec{B}\cdot d\vec{S} &=&%
\frac{e}{hc}\oint\limits_{\partial S}A\cdot dl=n, \\
\oint\limits_{\partial S}A\cdot dl &=&n\phi _{o},
\end{eqnarray*}%
where $\phi _{o}=\frac{hc}{e}$ is the quantum flux. This is reminiscent \ of
Eq. (\ref{b-s_version}) for the Landau level degeneracy values.

In the next section, we will discuss the B-S quantization condition in
coherent state (CS) formulation of quantum mechanics. This will introduce us
to the Berry phase and Berry connection in quantum field theory \cite{goff},
specifically non-Abelian gauge theory of elementary particles, which is
still an active field of research. For example, in effective quantum field
theory, the Berry phase is equivalent to Wess-Zumino-Witten action for the
sigma model \cite{hsin}.

\section{B-S quantization in CS formulation of quantum mechanics}

First we will treat the iconic harmonic oscillator in terms of ladder
operators. This is the precursor to the CS formulation of quantum physics.

\subsection{Harmonic oscillator}

The Hamiltonian is,

\begin{equation*}
H=\frac{p^{2}}{2m}+\frac{1}{2}\kappa q^{2},
\end{equation*}%
and the equations of motion are%
\begin{equation*}
\dot{q}=\frac{\partial H}{\partial p}=\frac{p}{m},\text{ \ }\dot{p}=-\frac{%
\partial H}{\partial q}=-\kappa q=-m\omega ^{2}q.
\end{equation*}%
We defined the ladder operators, i.e., annihilation and creation operators, $%
\alpha $\ and $\alpha ^{\dagger }$ as 
\begin{eqnarray*}
\alpha &=&\sqrt{\frac{m\omega }{2\hbar }}\left( q+\frac{ip}{m\omega }\right)
, \\
\alpha ^{\dagger } &=&\sqrt{\frac{m\omega }{2\hbar }}\left( q-\frac{ip}{%
m\omega }\right) .
\end{eqnarray*}%
In terms of this operators, we have,%
\begin{eqnarray*}
H &=&\left( \frac{1}{2}\kappa q^{2}+\frac{p^{2}}{2m}\right) \\
&=&\hbar \omega \left( \phi ^{\dagger }\phi \right) .
\end{eqnarray*}

\subsubsection{B-S quantization for harmonic oscillator}

We will try to implement the B-S quantization condition by using the
momentum and position operators in terms of the ladder operators and their
eigenvalues.

\begin{eqnarray*}
Q &=&\sqrt{\frac{\hbar }{2m\omega }}\left( \alpha +\alpha ^{\dagger }\right)
, \\
P &=&-i\sqrt{\frac{\hbar m\omega }{2}}\left( \alpha -\alpha ^{\dagger
}\right) ,
\end{eqnarray*}%
\begin{eqnarray*}
\left[ Q,P\right] &=&i\hbar , \\
\left[ \alpha ,\alpha ^{\dagger }\right] &=&1.
\end{eqnarray*}%
The B-S quantization now simply reads%
\begin{eqnarray}
\frac{1}{2\pi \hbar }\oint PdQ &=&\frac{-i}{2\pi \hbar }\oint \left( \sqrt{%
\frac{\hbar m\omega }{2}}\left( \alpha -\alpha ^{\dagger }\right) \right) \
d\left( \sqrt{\frac{\hbar }{2m\omega }}\left( \alpha +\alpha ^{\dagger
}\right) \right)  \notag \\
&=&\frac{-i}{2\pi \hbar }\frac{\hbar }{2}\oint \left( \alpha d\alpha -\alpha
^{\dagger }da^{\dagger }\right) +\left( ada^{\dagger }-a^{\dagger }da\right)
\notag \\
&=&n.  \label{BSoscillator}
\end{eqnarray}%
Although apparently the first term contributes to zero being exact integral
taken around the contour, there is actually an arbitrariness in the
constants of integration besides zeros. Thus, the first term involving $%
\alpha d\alpha -\alpha ^{\dagger }da^{\dagger }$ cannot be dismissed
immediately as zero, since it has have arbitrary constants of integration.

But taking advantage of the arbitrariness in constants of integration, we
may also conceive the following,%
\begin{eqnarray*}
\int \left( \alpha d\alpha -\alpha ^{\dagger }da^{\dagger }\right) &=&\int
\left( \left( \frac{\alpha ^{2}}{2}-i\gamma \right) -\left( \frac{\left(
\alpha ^{\dagger }\right) ^{2}}{2}+i\beta \right) \right) \\
&=&\int \left( \frac{\alpha ^{2}}{2}-\frac{\left( \alpha ^{\dagger }\right)
^{2}}{2}\right) -i\left( \gamma +\beta \right) \int d\phi =-i\left( \gamma
+\beta \right) 2\pi .
\end{eqnarray*}%
We may choose, $\left( \gamma +\beta \right) =1$. Then%
\begin{equation*}
\int \left( \alpha d\alpha -\alpha ^{\dagger }da^{\dagger }\right) =-i2\pi
=-i\frac{4\pi }{2}.
\end{equation*}%
Hence, we may have two quantization of the conjugate quantum-fields,%
\begin{equation*}
\frac{i}{4\pi }\left[ \int \left( a^{\dagger }da-ada^{\dagger }\right) %
\right] =n\text{,}
\end{equation*}%
and a possible quantization, by tailoring the constant of integration, with
hindsight of the zero-point energy of quantum harmonic oscillator as%
\begin{equation}
\frac{i}{4\pi }\left[ \int \left( a^{\dagger }da-ada^{\dagger }\right) %
\right] =n+\frac{1}{2}.  \label{zero_point}
\end{equation}%
Thus, the B-S quantization leads us to the correct positively definite
quantum energy levels of a harmonic oscillator,%
\begin{equation*}
E_{n}=\hbar \omega \left( n+\frac{1}{2}\right) ,
\end{equation*}%
which includes the zero-point energy, without actually solving for the
harmonic oscillator eigenfunctions.

\subsection{Coherent states and B-S quantization}

Coherent state is defined to be the \textit{right} eigenstate of the
annihilation operator $\hat{a}$,%
\begin{equation}
\hat{a}\ \left\vert \alpha \right\rangle =\alpha \ \left\vert \alpha
\right\rangle .  \label{ann-eigen-eq}
\end{equation}%
Since $\hat{a}$ is non-Hermitian, $\alpha =|\alpha |\ e^{i\theta }$ is
complex, $|\alpha |$ and $\theta $ are the amplitude and phase of the
eigenvalue, $\alpha $. The conjugate state $\left\langle \alpha \right\vert $
is the \textit{left} eigenstate of the creation operator $\hat{a}^{\dagger }$%
, this follows by taking the Hermitian conjugate of Eq. (\ref{ann-eigen-eq}%
). The state $\left\vert \alpha \right\rangle $ is the so-called coherent
state. The usefulness of coherent states is that they form a basis for the
representation of other states. Coherent states can never be made
orthogonal, although for well-separated eigenvalues $\alpha $, they can be
made approximately orthogonal \cite{buotbook}. Moreover, the set of coherent
states is overcomplete, in the sense that the set of coherent states form a
basis but are not linearly independent, i.e., they are expressible in terms
of each other. The nice thing is that the complex eigenvalue $\alpha $ is
labeled by the classical (average) values of position and momentum in the
following sense,%
\begin{equation*}
\left\langle \alpha \right\vert \hat{a}\left\vert \alpha \right\rangle
=\alpha =\sqrt{\frac{m\omega }{2\hbar }}\left\langle \alpha \right\vert
\left( Q+\frac{i}{m\omega }P\right) \left\vert \alpha \right\rangle ,
\end{equation*}%
\begin{equation}
\left\langle \alpha \right\vert Q\left\vert \alpha \right\rangle =\sqrt{%
\frac{2\hbar }{m\omega }}\func{Re}\alpha ,  \label{class_q-eq}
\end{equation}%
\begin{equation}
\left\langle \alpha \right\vert P\left\vert \alpha \right\rangle =\sqrt{%
2\hbar m\omega }\func{Im}\alpha ,  \label{class_p-eq}
\end{equation}%
\begin{eqnarray}
\func{Re}\alpha &=&\frac{1}{\sqrt{2}}\left\langle \alpha \right\vert \tilde{Q%
}\left\vert \alpha \right\rangle ,  \notag \\
\func{Im}\alpha &=&\frac{1}{\sqrt{2}}\left\langle \alpha \right\vert \tilde{P%
}\left\vert \alpha \right\rangle ,  \label{classical}
\end{eqnarray}%
\begin{eqnarray*}
2\hbar \func{Im}\alpha \func{Re}\alpha &=&\frac{1}{2}\left\langle \alpha
\right\vert \tilde{P}\left\vert \alpha \right\rangle \left\langle \alpha
\right\vert \tilde{Q}\left\vert \alpha \right\rangle \\
&=&\left\langle \alpha \right\vert P\left\vert \alpha \right\rangle
\left\langle \alpha \right\vert Q\left\vert \alpha \right\rangle ,
\end{eqnarray*}%
where $\tilde{Q}$ and $\tilde{P}$ are the scaled canonical operators given
by, 
\begin{eqnarray}
\tilde{Q} &=&\sqrt{\frac{m\omega }{\hbar }}Q,  \notag \\
\tilde{P} &=&\sqrt{\frac{1}{\hbar m\omega }}P.  \label{scaled_mom_pos2}
\end{eqnarray}%
Indeed, we have%
\begin{eqnarray}
\alpha &=&\frac{1}{\sqrt{2}}\left( \left\langle \alpha \right\vert \tilde{Q}%
\left\vert \alpha \right\rangle +i\left\langle \alpha \right\vert \tilde{P}%
\left\vert \alpha \right\rangle \right)  \notag \\
&=&\left\langle \alpha \right\vert \hat{a}\left\vert \alpha \right\rangle .
\label{alpha-P-Q}
\end{eqnarray}

The counting of $\alpha $ states can be derived from the counting of states
in $\left( p_{c},q_{c}\right) $ phase-space, where $q_{c}=\left\langle
\alpha \right\vert Q\left\vert \alpha \right\rangle $ and $%
p_{c}=\left\langle \alpha \right\vert P\left\vert \alpha \right\rangle $,
namely,%
\begin{eqnarray*}
\frac{1}{2\pi \hbar }dq_{c}dp_{c} &=&\frac{1}{\pi }d\func{Re}\alpha \ d\func{%
Im}\alpha \\
&\Rightarrow &\frac{1}{\pi }d\alpha ^{2}\text{ (abbreviated).}
\end{eqnarray*}%
Since $\left\langle \alpha \right\vert P\left\vert \alpha \right\rangle $
appears as the Berry connection, we proceed to apply the B-S quantization as
follows%
\begin{eqnarray}
\frac{2\hbar }{2\pi \hbar }\oint \func{Im}\alpha \ d\func{Re}\alpha &=&n, 
\notag \\
\frac{-i}{4\pi }\oint \left( \alpha -\alpha ^{\dagger }\right) \ d\left(
\alpha +\alpha ^{\dagger }\right) &=&n,  \notag \\
\frac{-i}{4\pi }\oint \left( \alpha d\alpha -\alpha ^{\dagger }da^{\dagger
}\right) +\frac{i}{4\pi }\oint \left( \alpha ^{\dagger }d\alpha -\alpha \
d\alpha ^{\dagger }\right) \ &=&n.  \label{BS_CS}
\end{eqnarray}%
Equation (\ref{BS_CS}) for the CS representation exactly reproduce Eq. (\ref%
{BSoscillator}) for the harmonic oscillator. The same considerations for the
first term of Eq. (\ref{BS_CS}) yields the same expression as Eq. (\ref%
{zero_point}).

In agreement with published results, the second term of Eq.(\ref{BS_CS})
appears as an approximate (semiclassical) result in the coherent state
representation, as obtained in a B-S quantization by Tochishita et al \cite%
{tochishita} written as Eq. (38) in their paper,

\begin{equation}
\frac{i\hbar }{2}\oint\limits_{C}\left( \xi ^{\ast }d\xi -\xi d\xi ^{\ast
}\right) =2\pi m\hbar .  \label{tochishita}
\end{equation}

\section{Other areas were B-S condition have been used}

There are several instances that the B-S condition have been used, which are
beyond the scope of this paper. For example, it has been used for
one-dimensional Bogoliubov-de Gennes Hamiltonian. It may also have relevance
to the so-called Fermionic oscillator in the $2$-state system and its
corresponding coherent state formulation. Its relevance and utility around
non-Abelian gauge theory is indeed not yet clear and lack focused
investigation.

In mathematics of geometric quantization, monodromy (and holonomy) of
completely integrable Hamiltonian systems makes use of B--S quantization
condition \cite{sansoneto}.

\subsection{B-S quantization in FQHE}

Recently, the B-S quantization has been employed \cite{jacak1,jacak2,jacak3}
in the study of FQHE which generalizes the composite fermion theory of Jain 
\cite{jain}. In the composite fermion theory, the concept of fractional
charge is central to the theory of the FQHE, thus the FQHE conductance
essentially follows the formula for IQHE with renormalized electron charge.
In the recent work of Jacak \cite{jacak1,jacak2,jacak3}, the B-S
quantization has been employed in his holonomic approach using multi-loop
orbits for incommensurate ratio of Wigner lattice constant to the magnetic
length, $\frac{a}{l_{B}}=\frac{p}{q}$. His holonomic approach has produce a
new hierarchy in FQHE which include the composite fermion heirarchy of Jain 
\cite{jain}, as well as the IQHE. Indeed it is the commensuration and
incommensuration of these two length scales that lead to the fractal
spectrum of Hofstadter butterfly \cite{hopsta} and Wannier Diophantine
equation \cite{wadio} for the gaps of fractal spectrum.

The work on FQHE is still a topic of active research. Indeed, the topic of
the interaction of magnetic field with matter has produced so many enigmatic
and intriguing results in the history of physics, to name a few, such as the
fractal spectrum of Hofstadter butterfly, IQHE, giant diamagnetism \cite%
{diamag}, and of course the ongoing active research in FQHE and fractal
spectrum. All these may just be a 'tip of the iceberg' in revealing the
fundamental understanding of nature which are pursued in so many diverse
fronts in both geometry, physics, and the sciences.

In the following section, we will discuss in more details a novel quantum
transport approach to the B-S quantization of the IQHE in electrical
conductivity. The physics of IQHE is well established, however from the
point of view of B-S quantization this is not fully appreciated. Moreover, a
fully kinetic approach has not been given, which is in stark contrast to
other approaches, such as the conventional Kubo current-current correlation
perturbative approach of equilibrium systems.

\section{B-S quantization in a novel kinetic approach to IQHE}

Here we present our new approach to integer quantum Hall effect which makes
use of quantum superfield lattice Weyl Transform nonequilibrium Green's
function (SFLWT-NEGF) formalism \cite{buotbook}. This formalism includes
nonequilibrium transport in superconductivity, as well as vacuum
fluctuations or zitterbewegung in conductance measurements \cite{iwasaki}.

The calculation of integer quantum Hall effect is another good example where
the IQHE conductance is directly proportional to the B-S quantization of Eq.
(\ref{eq1}), i.e.,%
\begin{equation}
\sigma _{xy}=\frac{e^{2}}{h}\sum\limits_{\alpha }\frac{\Delta \phi _{total}}{%
2\pi }=\frac{e^{2}}{h}\sum\limits_{\alpha }n_{\alpha },  \label{QHEquantum}
\end{equation}%
where B-S condition enters in the following simple form,%
\begin{eqnarray*}
\frac{\Delta \phi _{total}}{2\pi } &=&\frac{1}{2\pi }\oint dk_{c}\ \left[
\left\langle \alpha ,\vec{k}\right\vert i\frac{\partial }{\partial k_{c}}%
\left\vert \alpha ,\vec{k}\right\rangle \right] _{contour} \\
&=&\frac{1}{2\pi }\oint \ \left( \nabla _{c}\phi \right) \ dk_{c}=n_{\alpha
},
\end{eqnarray*}%
where $\frac{e^{2}}{h}$ is the \textit{quantum conductance}. Again $\Delta
\phi _{total}$ is identified as the B-S contour integral undergoing
quantization. This will be derived in this section since our new quantum
transport approach to IQHE is not well known.

It is also worth mentioning that in mesoscopic closed circuits with
ballistic conducting channel and perfectly conducting leads, the conductance
is simply equal to $\frac{e^{2}}{h}$, the \textit{quantum conductance per
electron spin, i.e., }$n_{\sigma }\equiv 1$\textit{\ }\cite{buotbook}\textit{%
. }

The following discussions are intended to derive Eq. (\ref{QHEquantum}) from
a kinetic quantum transport employing phase-space quantum distribution or
Wigner distribution function.

\subsection{The Wigner distribution and density matrix operator}

If we write the second quantized operator for the one-particle $\left( \vec{p%
},\vec{q};E,t\right) $-phase space distribution function as%
\begin{equation}
\hat{f}_{\lambda \lambda \sigma \sigma }\left( \vec{p},\vec{q};E,t\right)
=\sum\limits_{v}e^{\frac{2i}{\hbar }p\cdot v}\psi _{\lambda \sigma
}^{\dagger }\left( q+v,t+\frac{\tau }{2}\right) \psi _{\lambda ^{\prime
}\sigma ^{\prime }}\left( q-v,t-\frac{\tau }{2}\right) \text{,}
\label{distop}
\end{equation}%
where $\lambda $ label the band index and $\sigma $ the spin index [here we
drop the Heisenberg representation subscripts $H$ for economy of indices],
then upon taking the average%
\begin{equation}
\left\langle \hat{f}_{\lambda \lambda ^{\prime }\sigma \sigma ^{\prime
}}\left( \vec{p},\vec{q};E,t\right) \right\rangle =\sum\limits_{v}e^{\frac{2i%
}{\hbar }p\cdot v}\left\langle \psi _{\lambda \sigma }^{\dagger }\left( \vec{%
q}+\vec{v},t+\frac{\tau }{2}\right) \psi _{\lambda ^{\prime }\sigma ^{\prime
}}\left( \vec{q}-\vec{v},t-\frac{\tau }{2}\right) \right\rangle \text{,}
\label{lwt_rho}
\end{equation}%
we obtain particle distribution function $\rho _{\lambda \lambda ^{\prime
}\sigma \sigma ^{\prime }}\left( p,q\right) ,$ a generalized Wigner
distribution function,%
\begin{equation*}
\rho _{\lambda \lambda ^{\prime }\sigma \sigma ^{\prime }}\left( p,q\right)
=\left\langle \hat{f}_{\lambda \lambda ^{\prime }\sigma \sigma ^{\prime
}}\left( p,q\right) \right\rangle \text{,}
\end{equation*}%
where we employ the four-dimensional notation: $p=\left( \vec{p},E\right) $
and $q=\left( \vec{q},t\right) $. Equation (\ref{lwt_rho}) is indeed the
lattice Weyl transform of the density matrix operator $\hat{\rho}$ as%
\begin{eqnarray*}
\rho _{\lambda \lambda ^{\prime }\sigma \sigma ^{\prime }}\left( p,q\right)
&=&\left\langle \hat{f}_{\lambda \lambda ^{\prime }\sigma \sigma ^{\prime
}}\left( p,q\right) \right\rangle \\
&=&\sum\limits_{v}\exp \left( \frac{2i}{\hbar }p\cdot v\right) \left\langle
q-v;\lambda ^{\prime },\sigma ^{\prime }\right\vert \ \hat{\rho}\ \left\vert
q+v;\lambda ,\sigma \right\rangle \text{,}
\end{eqnarray*}%
where the RHS is the lattice Weyl transform (LWT) of the density matrix
operator, which is identical to the LWT of $-iG^{<}\left( 1,2\right) $,
where $-iG^{<}\left( 1,2\right) =Tr\left[ \rho _{H}\left( \psi _{H}^{\dagger
}\left( 2\right) \right) \psi _{H}\left( 1\right) \right] =\left\langle
\left( \psi _{H}^{\dagger }\left( 2\right) \right) \psi _{H}\left( 1\right)
\right\rangle .$The indices $1$ and $2$ subsume all space-time indices and
other quantum-label indices. $\psi _{H}\left( 1\right) $ and $\psi
_{H}^{\dagger }\left( 2\right) $ are the particle annihilation and creation
operators in the Heisenberg representation, respectively.

The expectation value of one-particle operator $\hat{A}$ can be calculated 
\textit{in phase-space} similar to the classical averages using a
distribution function,%
\begin{equation*}
Tr\left( \hat{\rho}\ \hat{A}\right) :=\left\langle \hat{A}\right\rangle
=\sum\limits_{p,q;\lambda \lambda ^{\prime }\sigma \sigma ^{\prime
}}A_{\lambda \lambda ^{\prime }\sigma \sigma ^{\prime }}\left( p,q\right)
\rho _{\lambda ^{\prime }\lambda \sigma ^{\prime }\sigma }\left( p,q\right) 
\text{,}
\end{equation*}%
clearly exhibiting the trace of binary operator product as a trace of the
product of their respective LWT's. This general observation is crucial in
most of the calculations that follows.

The Wigner distribution function $f_{W}\left( \vec{p},\vec{q},t\right) $ is
given by integrating out the energy variable,%
\begin{equation*}
f_{W}\left( \vec{p},\vec{q},t\right) =\frac{1}{2\pi }\int dE\left(
-iG^{<}\left( \vec{p},\vec{q};E,t\right) \right) \text{.}
\end{equation*}%
We further note that 
\begin{eqnarray}
\rho \left( t\right) &=&e^{-\frac{i}{\hbar }Ht}\rho \left( 0\right) e^{\frac{%
i}{\hbar }Ht}  \notag \\
&=&U\left( t\right) \rho \left( 0\right) U^{\dagger }\left( t\right) \text{,}
\label{timedepend}
\end{eqnarray}%
provides the major time dependence in the transport equation that follows.

\subsubsection{\textbf{Spatio-temporal translation operators: action
principle}}

We define translation operator in space and time, which commute, i.e., $%
\left[ T\left( q\right) ,T\left( t\right) \right] =0$, as 
\begin{equation}
T\left( q\right) T\left( t\right) =:\exp \frac{i}{\hbar }\left( \hat{P}\cdot
q-{\LARGE \hat{\varepsilon}t}\right) ,  \label{translation_op_ST}
\end{equation}%
where $\hat{P}$ is the momentum operator, $-i\hbar \nabla _{\mathbf{q}}\psi =%
\hat{P}\psi $, and ${\LARGE \hat{\varepsilon}}$ is the energy operator,
explicitly given by ${\LARGE \hat{\varepsilon}}=i\hbar \frac{\partial }{%
\partial t}$ since in the Schr\"{o}dinger equation, $i\hbar \frac{\partial }{%
\partial t}\psi =\hat{H}\psi $ . The phase of the translation operator
mirrors that of the Lagrangian, $\mathcal{L}$, in Hamilton classical
mechanics, namely, 
\begin{equation*}
\int dt\ \mathcal{L}=\int dt\sum\limits_{q_{i}}p_{i}\dot{q}_{i}-H,
\end{equation*}%
where the stationary action gives the classical equation of motion.

Commutation properties of the time and space translation operators, $\hat{T}%
\left( t\right) $ and $\hat{T}\left( \vec{q}\right) $, are given by,%
\begin{equation}
\frac{\partial \hat{T}\left( t\right) }{\partial \vec{q}}=\frac{i}{\hbar }%
\left[ \mathcal{\vec{K}},\hat{T}\left( t\right) \right] =\left( \frac{i}{%
\hbar }e\vec{F}t\right) \ \hat{T}\left( t\right) ,  \label{eq3}
\end{equation}%
with respect to displacement in lattice position, and%
\begin{equation}
\frac{\partial \hat{T}\left( \vec{q}\right) }{\partial t}=\frac{i}{\hbar }%
\left[ \mathcal{H},\hat{T}\left( \vec{q}\right) \right] =\left( \frac{i}{%
\hbar }e\vec{F}.\vec{q}\right) \ \hat{T}\left( \vec{q}\right) \text{,}
\label{eq2}
\end{equation}%
with respect to displacement in time. In Eq. (\ref{eq3}), the canonical
momentum $\mathcal{\vec{K}}$ is given by Eq. (\ref{kappa}) below in a
self-consistent manner. These results suggest that in the presence of
electric field, gauge invariant quantities that are \textit{displaced} in
space and time acquires Peierls phase factors \cite{bj}. For example, a
nonlocal matrix element acquires a generalized Peierls phase factor as 
\begin{equation}
\left\langle \vec{q}_{1},t_{1}\right\vert \hat{H}^{\left( 1\right)
}\left\vert \vec{q}_{2},t_{2}\right\rangle \Longrightarrow e^{-i\frac{e}{%
\hbar }\vec{F}t\cdot \left( \vec{q}_{1}-\vec{q}_{2}\right) }e^{-i\frac{e}{%
\hbar }\vec{F}\cdot \vec{q}\left( t_{1}-t_{2}\right) }H^{\left( 1\right)
}\left( \vec{q}_{1}-\vec{q}_{2},t_{1}-t_{2}\right) \text{,}
\label{peierlsPhase}
\end{equation}%
where%
\begin{equation*}
\vec{q}=\frac{1}{2}\left( \vec{q}_{1}+\vec{q}_{2}\right) \text{, \ \ }t=%
\frac{1}{2}\left( t_{1}+t_{2}\right) \text{.}
\end{equation*}%
Using the four dimensional notation: $p=\left( \vec{p},E\right) $ and $%
q=\left( \vec{q},t\right) $, the Weyl transform $A\left( p,q\right) $ of any
operator $\mathbf{\hat{A}}$ is defined by%
\begin{eqnarray}
A_{\lambda \lambda ^{\prime }}\left( p.q\right) &=&\sum\limits_{v}e^{\left( 
\frac{2i}{\hbar }\right) p\cdot v}\left\langle q-v,\lambda \right\vert 
\mathbf{\hat{A}}\left\vert q+v,\lambda ^{\prime }\right\rangle  \label{LWT}
\\
&=&\sum\limits_{u}e^{\left( \frac{2i}{\hbar }\right) q\cdot u}\left\langle
p+u,\lambda \right\vert \mathbf{\hat{A}}\left\vert p-u,\lambda ^{\prime
}\right\rangle \text{,}  \label{LWT1}
\end{eqnarray}%
where $\lambda $ and $\lambda ^{\prime }$ stands for other discrete quantum
numbers. Viewed as a transformation of a matrix, we see that the Weyl
transform of the matrix $\left\langle q^{\prime },\lambda ^{\prime
}\right\vert \mathbf{\hat{A}}\left\vert q^{\prime \prime },\lambda
^{^{\prime \prime }}\right\rangle $ is given by Eq. (\ref{LWT}) and the
lattice Weyl transform of $\left\langle p^{\prime },\lambda ^{\prime
}\right\vert \mathbf{\hat{A}}\left\vert p^{\prime \prime },\lambda ^{\prime
\prime }\right\rangle $ is given by Eq. (\ref{LWT1}). Denoting the operation
of taking the lattice Weyl transform by the symbol $\mathcal{W}$ then it is
easy to see that the lattice Weyl transform of 
\begin{eqnarray}
\mathcal{W}\left( \frac{\partial }{\partial q^{\prime }}+\frac{\partial }{%
\partial q^{\prime \prime }}\right) \left\langle q^{\prime },\lambda
^{\prime }\right\vert \mathbf{\hat{A}}\left\vert q^{\prime \prime },\lambda
^{^{\prime \prime }}\right\rangle &=&\frac{\partial }{\partial q}%
\sum\limits_{v}e^{\left( \frac{2i}{\hbar }\right) p\cdot v}\left\langle
q-v,\lambda ^{\prime }\right\vert \mathbf{\hat{A}}\left\vert q+v,\lambda
^{\prime \prime }\right\rangle  \notag \\
&=&\frac{\partial }{\partial q}A_{\lambda ^{\prime }\lambda ^{\prime \prime
}}\left( p.q\right) \text{.}  \label{sumq}
\end{eqnarray}%
Similarly 
\begin{eqnarray}
\mathcal{W}\left( \frac{\partial }{\partial p^{\prime }}+\frac{\partial }{%
\partial p^{\prime \prime }}\right) \left\langle p^{\prime },\lambda
^{\prime }\right\vert \mathbf{\hat{A}}\left\vert p^{\prime \prime },\lambda
^{^{\prime \prime }}\right\rangle &=&\frac{\partial }{\partial p}%
\sum\limits_{u}e^{\left( \frac{2i}{\hbar }\right) q\cdot u}\left\langle
p+u,\lambda \right\vert \mathbf{\hat{A}}\left\vert p-u,\lambda ^{\prime
}\right\rangle  \notag \\
&=&\frac{\partial }{\partial p}A_{\lambda ^{\prime }\lambda ^{\prime \prime
}}\left( p.q\right) \text{.}  \label{sump}
\end{eqnarray}%
Note that the derivatives on the LHS of Eqs. (\ref{sumq}) and (\ref{sump})
operate only on the wavefunction or state vectors.

Writing Eq. (\ref{LWT}) explicitly, we have%
\begin{equation}
A_{\lambda \lambda ^{\prime }}\left( \vec{p}.\vec{q};E,t\right)
=\sum\limits_{\vec{v};\tau }e^{\left( \frac{2i}{\hbar }\right) \vec{p}\cdot 
\vec{v}}e^{\left( \frac{i}{\hbar }\right) E\tau }\left\langle \vec{q}-\vec{v}%
;t-\frac{\tau }{2},\lambda \right\vert \mathbf{\hat{A}}\left\vert \vec{q}+%
\vec{v};t+\frac{\tau }{2},\lambda ^{\prime }\right\rangle \text{.}
\label{LWT2}
\end{equation}%
Using the form of 'nonlocal' matrix elements in Eq. (\ref{peierlsPhase}), we
have%
\begin{equation*}
\left\langle \vec{q}-\vec{v};t-\frac{\tau }{2},\lambda \right\vert \mathbf{%
\hat{A}}\left\vert \vec{q}+\vec{v};t+\frac{\tau }{2},\lambda ^{\prime
}\right\rangle =e^{-i\frac{e}{\hbar }\vec{F}t\cdot \left( \vec{q}_{1}-\vec{q}%
_{2}\right) }e^{-i\frac{e}{\hbar }\vec{F}\cdot \vec{q}\left(
t_{1}-t_{2}\right) }A\left( \vec{q}_{1}-\vec{q}_{2},t_{1}-t_{2}\right) .
\end{equation*}%
Thus%
\begin{eqnarray*}
A_{\lambda \lambda ^{\prime }}\left( \vec{p},\vec{q};E,t\right)
&=&\sum\limits_{\vec{v};\tau }e^{\left( \frac{2i}{\hbar }\right) \vec{p}%
\cdot \vec{v}}e^{\left( \frac{i}{\hbar }\right) E\tau }e^{i\frac{e}{\hbar }%
\vec{F}t\cdot \left( 2v\right) }e^{i\frac{e}{\hbar }\vec{F}\cdot \vec{q}\tau
}A_{\lambda \lambda ^{\prime }}\left( 2\vec{v},\tau \right) \\
&=&\sum\limits_{\vec{v};\tau }e^{\left( \frac{2i}{\hbar }\right) \left( \vec{%
p}+e\vec{F}t\right) \cdot \vec{v}}e^{\left( \frac{i}{\hbar }\right) \left(
E+e\vec{F}\cdot \vec{q}\right) \tau }A_{\lambda \lambda ^{\prime }}\left( 2%
\vec{v},\tau \right) \\
&=&A_{\lambda \lambda ^{\prime }}\left( \mathcal{\vec{K}};\mathcal{E}\right) 
\text{.}
\end{eqnarray*}%
Hence the expected dynamical variables in the phase space including the time
variable occur in $\mathcal{\vec{K}}$ and $\mathcal{E}$. Therefore, besides\
the crystal momentum varying in time as 
\begin{equation}
\mathcal{\vec{K}}=\vec{p}+e\vec{F}t\text{,}  \label{kappa}
\end{equation}%
consistent with Eq. (\ref{eq3}), the energy variable varies with $\vec{q}$
as 
\begin{eqnarray}
\mathcal{E} &=&E+e\vec{F}\cdot \vec{q}  \notag \\
&=&E_{o}\left( \mathcal{\vec{K}}\right) +e\vec{F}\cdot \vec{q}\text{.}
\label{shiftedE}
\end{eqnarray}%
In effect we have unified the use of \textit{scalar potential} in $\mathcal{E%
}$ and \textit{vector potential} in $\mathcal{\vec{K}}$ for a system under
uniform electric fields. The LWT of the effective or renormalized lattice
Hamiltonian $\mathcal{H}_{eff}\leftrightarrows H\left( \vec{p},\vec{q}%
;E,t\right) $ can therefore be analyzed on $\left( \mathcal{\vec{K}},%
\mathcal{E}\right) $-space as%
\begin{eqnarray*}
H\left( \vec{p},\vec{q};E,t\right) &=&H\left( \mathcal{\vec{K}},\mathcal{E}%
\right) \\
&=&H_{o}\left( \mathcal{\vec{K}}\right) +e\vec{F}\cdot \vec{q}\text{.}
\end{eqnarray*}%
The last line is by virtue of Eq. (\ref{shiftedE}). And all gauge invariant
quantities are functions of $\left( \mathcal{\vec{K}},\mathcal{E}\right) ,$
such as the \textit{electric Bloch function\cite{bj}} or Houston
wavefunction [\cite{zener}] and \textit{electric Wannier function}, i.e.,
the electric-field dependent generalization of Wannier function by
essentially endowing it with a Peierls phase factor. Observe that in the
absence of the electric field, the dependence $\left( \mathcal{\vec{K}},%
\mathcal{E}\right) \Longrightarrow _{F\Longrightarrow 0}\left( \vec{p},\hbar
\omega \right) $ indicating a translationally symmetric system at stationary
state where $\omega $ is the frequency in the absence of uniform electric
field.

The Weyl transform of a commutator,%
\begin{equation*}
\mathcal{W}\left[ H,G^{<}\right] =\sin \Lambda \text{,}
\end{equation*}%
where $\Lambda $ is the Poisson bracket operator,%
\begin{eqnarray}
\Lambda &=&\frac{\hbar }{2}\left[ \frac{\partial ^{\left( a\right) }}{%
\partial t}\frac{\partial ^{\left( b\right) }}{\partial \mathcal{E}}-\frac{%
\partial ^{\left( a\right) }}{\partial \mathcal{E}}\frac{\partial ^{\left(
b\right) }}{\partial t}\right] \text{ }  \notag \\
&\Longrightarrow &\frac{\hbar }{2}\frac{\partial \mathcal{\vec{K}}}{\partial
t}\cdot \left[ \frac{\partial ^{\left( a\right) }}{\partial \mathcal{\vec{K}}%
}\frac{\partial ^{\left( b\right) }}{\partial \mathcal{E}}-\frac{\partial
^{\left( a\right) }}{\partial \mathcal{E}}\frac{\partial ^{\left( b\right) }%
}{\partial \mathcal{\vec{K}}}\right]  \notag \\
&=&\frac{\hbar }{2}e\vec{F}\cdot \left[ \frac{\partial ^{\left( a\right) }}{%
\partial \mathcal{\vec{K}}}\frac{\partial ^{\left( b\right) }}{\partial 
\mathcal{E}}-\frac{\partial ^{\left( a\right) }}{\partial \mathcal{E}}\frac{%
\partial ^{\left( b\right) }}{\partial \mathcal{\vec{K}}}\right] \text{, }
\label{poissonBracket}
\end{eqnarray}%
on $\left( \mathcal{\vec{K}},\mathcal{E}\right) $-phase space.

\subsection{SFLWT-NEGF transport equation}

The nonequilibrium quantum superfield transport equation for interacting
Bloch electrons under a uniform electric field has been derived in Sec. $VI$
of Buot and Jensen paper \cite{bj}. In the absence of superconducting
behavior and Zitterbewegung, the SFLWT-NEGF phase-space transport equation
reads%
\begin{eqnarray}
\frac{\partial }{\partial t}G^{<}\left( \vec{p},\vec{q};E,t\right) &=&\frac{2%
}{\hbar }\sin \hat{\Lambda}\left\{ H\left( p,q\right) G^{<}\left( p,q\right)
+\Sigma ^{<}\left( p,q\right) \func{Re}G^{r}\left( p,q\right) \right\} 
\notag \\
&&+\frac{1}{\hbar }\cos \hat{\Lambda}\left\{ \Sigma ^{<}\left( p,q\right)
A\left( p,q\right) -\Gamma \left( p,q\right) G^{<}\left( p,q\right) \right\} 
\text{.}  \label{eq105}
\end{eqnarray}%
If we expand Eq. (\ref{eq105}) to first order in the gradient, i.e., $\sin
\Lambda \simeq \Lambda ,$we obtain%
\begin{eqnarray}
\frac{\partial }{\partial t}G^{<}\left( p,q\right) &=&-e\vec{F}\cdot \left\{ 
\frac{\partial }{\partial \partial \mathcal{E}}E_{\alpha }\left( \vec{K}%
\right) +\frac{\partial \func{Re}\Sigma ^{r}\left( p,q\right) }{\partial 
\mathcal{E}}\right\} \frac{\partial }{\partial \mathcal{\vec{K}}}G^{<}\left(
p,q\right)  \notag \\
&&+e\vec{F}\cdot \left\{ \frac{\partial }{\partial \mathcal{\vec{K}}}%
E_{\alpha }\left( \vec{K}\right) +\frac{\partial \func{Re}\Sigma ^{r}\left(
p,q\right) }{\partial \mathcal{\vec{K}}}\right\} \frac{\partial }{\partial 
\mathcal{E}}G^{<}\left( p,q\right)  \notag \\
&&-e\vec{F}\cdot \left\{ \frac{\partial \Sigma ^{<}\left( p,q\right) }{%
\partial \mathcal{E}}\frac{\partial \func{Re}G^{r}\left( p,q\right) }{%
\partial \mathcal{\vec{K}}}\right\} +e\vec{F}\cdot \left\{ \frac{\partial
\Sigma ^{<}\left( p,q\right) }{\partial \mathcal{\vec{K}}}\frac{\partial 
\func{Re}G^{r}\left( p,q\right) }{\partial \mathcal{E}}\right\}  \notag \\
&&+\frac{1}{\hbar }\left\{ \Sigma ^{<}\left( p,q\right) A\left( p,q\right)
-\Gamma \left( p,q\right) G^{<}\left( p,q\right) \right\} \text{,}
\label{eq_grad}
\end{eqnarray}%
where $G^{r}\left( p,q\right) $ is the LWT of the retarded Green's function, 
$A\left( p,q\right) $ is the spectral function, and $\Gamma \left(
p,q\right) $ is the corresponding scattering rate.

\subsubsection{\textbf{Ballistic transport and diffusion}}

We wiill simplify Eq. (\ref{eq_grad}) by neglecting the self-energies, i.e.,
we limit to non-interacting particles. Then we have the following simplified
quantum transport equation,%
\begin{eqnarray}
\frac{\partial }{\partial t}G^{<}\left( p,q\right) &=&-e\vec{F}\cdot \frac{%
\partial }{\partial \mathcal{E}}E_{\alpha }\left( \vec{K}\right) \frac{%
\partial }{\partial \mathcal{\vec{K}}}G^{<}\left( p,q\right)  \notag \\
&&+e\vec{F}\cdot \frac{\partial }{\partial \mathcal{\vec{K}}}E_{\alpha
}\left( \vec{K}\right) \frac{\partial }{\partial \mathcal{E}}G^{<}\left(
p,q\right) ,  \notag
\end{eqnarray}%
which can be written in terms of the Poisson bracket of Eq. (\ref%
{poissonBracket}) as%
\begin{equation}
\frac{\partial }{\partial t}G^{<}\left( \mathcal{\vec{K}},\mathcal{E}\right)
=\frac{2}{\hbar }\frac{\hbar }{2}e\vec{F}\cdot \left[ \frac{\partial
^{\left( a\right) }}{\partial \mathcal{\vec{K}}}\frac{\partial ^{\left(
b\right) }}{\partial \mathcal{E}}-\frac{\partial ^{\left( a\right) }}{%
\partial \mathcal{E}}\frac{\partial ^{\left( b\right) }}{\partial \mathcal{%
\vec{K}}}\right] H^{\left( a\right) }\left( \mathcal{\vec{K}},\mathcal{E}%
\right) G^{<\left( b\right) }\left( \mathcal{\vec{K}},\mathcal{E}\right) 
\text{.}  \label{non_interact}
\end{equation}%
Therefore%
\begin{equation*}
G^{<}\left( \mathcal{\vec{K}},\mathcal{E}\right) =e\vec{F}\cdot \int dt\ %
\left[ \frac{\partial ^{\left( a\right) }}{\partial \mathcal{\vec{K}}}\frac{%
\partial ^{\left( b\right) }}{\partial \mathcal{E}}-\frac{\partial ^{\left(
a\right) }}{\partial \mathcal{E}}\frac{\partial ^{\left( b\right) }}{%
\partial \mathcal{\vec{K}}}\right] H^{\left( a\right) }\left( \mathcal{\vec{K%
}},\mathcal{E}\right) G^{<\left( b\right) }\left( \mathcal{\vec{K}},\mathcal{%
E}\right) \text{.}
\end{equation*}%
Assuming that the electric field is in the $x$-direction. Then%
\begin{equation*}
G^{<}\left( \mathcal{\vec{K}},\mathcal{E}\right) =e\left\vert \vec{F}%
\right\vert \int dt\ \left[ \frac{\partial ^{\left( a\right) }}{\partial 
\mathcal{\vec{K}}_{x}}\frac{\partial ^{\left( b\right) }}{\partial \mathcal{E%
}}-\frac{\partial ^{\left( a\right) }}{\partial \mathcal{E}}\frac{\partial
^{\left( b\right) }}{\partial \mathcal{\vec{K}}_{x}}\right] H^{\left(
a\right) }\left( \mathcal{\vec{K}},\mathcal{E}\right) G^{<\left( b\right)
}\left( \mathcal{\vec{K}},\mathcal{E}\right) \text{.}
\end{equation*}%
The Hall current in the $y$-direction is thus determined by the following
equation, 
\begin{eqnarray}
&&\frac{a^{2}}{\left( 2\pi \hbar \right) ^{2}}\int \int d\mathcal{\vec{K}}%
_{x}d\mathcal{\vec{K}}_{y}\left( \frac{e}{a^{2}}\frac{\partial \mathcal{E}}{%
\partial \mathcal{\vec{K}}_{y}}\right) \left( -iG^{<}\left( \mathcal{\vec{K}}%
,\mathcal{E}\right) \right)  \notag \\
&=&e^{2}\left\vert \vec{F}\right\vert \frac{1}{\left( 2\pi \hbar \right) ^{2}%
}\int \int \int \ d\mathcal{\vec{K}}_{x}d\mathcal{\vec{K}}_{y}\ dt\frac{%
\partial \mathcal{E}}{\partial \mathcal{\vec{K}}_{y}}\left[ \frac{\partial
^{\left( a\right) }}{\partial \mathcal{\vec{K}}_{x}}\frac{\partial ^{\left(
b\right) }}{\partial \mathcal{E}}-\frac{\partial ^{\left( a\right) }}{%
\partial \mathcal{E}}\frac{\partial ^{\left( b\right) }}{\partial \mathcal{%
\vec{K}}_{x}}\right]  \notag \\
&&\times H^{\left( a\right) }\left( \mathcal{\vec{K}},\mathcal{E}\right)
\left( -iG^{<\left( b\right) }\left( \mathcal{\vec{K}},\mathcal{E}\right)
\right)  \notag \\
&=&e^{2}\left\vert \vec{F}\right\vert \frac{1}{\left( 2\pi \hbar \right) ^{2}%
}\int \int \int d\mathcal{\vec{K}}_{x}d\mathcal{\vec{K}}_{y}dt\ \left[ \frac{%
\partial ^{\left( a\right) }}{\partial \mathcal{\vec{K}}_{x}}\frac{\partial
^{\left( b\right) }}{\partial \mathcal{\vec{K}}_{y}}-\frac{\partial ^{\left(
a\right) }}{\partial \mathcal{\vec{K}}_{y}}\frac{\partial ^{\left( b\right) }%
}{\partial \mathcal{\vec{K}}_{x}}\right]  \notag \\
&&\times H^{\left( a\right) }\left( \mathcal{\vec{K}},\mathcal{E}\right)
\left( -iG^{<\left( b\right) }\left( \mathcal{\vec{K}},\mathcal{E}\right)
\right) \text{.}  \label{eqKE}
\end{eqnarray}%
If we are only interested in linear response we may consider all the
quantities in the integrand to be of zero-order in the electric field,
although this is not necessary if we allow for very weak electric field
leading to time dependence being dominated by the time dependence of the
density matrix, as we shall see in what follows.

\subsubsection{\textbf{Topological invariant in }$\left( p,q;E,t\right) $%
\textbf{-space}}

From Eq. (\ref{eqKE}), we claim that the quantized Hall conductivity is
given by 
\begin{equation}
\sigma _{yx}=\left( \frac{e^{2}}{h}\right) \frac{1}{\left( 2\pi \hbar
\right) }\int \int \int d\mathcal{\vec{K}}_{x}d\mathcal{\vec{K}}_{y}dt\ %
\left[ \frac{\partial ^{\left( a\right) }}{\partial \mathcal{\vec{K}}_{x}}%
\frac{\partial ^{\left( b\right) }}{\partial \mathcal{\vec{K}}_{y}}-\frac{%
\partial ^{\left( a\right) }}{\partial \mathcal{\vec{K}}_{y}}\frac{\partial
^{\left( b\right) }}{\partial \mathcal{\vec{K}}_{x}}\right] H^{\left(
a\right) }\left( \mathcal{\vec{K}},\mathcal{E}\right) \left( -iG^{<\left(
b\right) }\left( \mathcal{\vec{K}},\mathcal{E}\right) \right) \text{,}
\label{hallcond}
\end{equation}%
and is quantized in units of $\frac{e^{2}}{h}$, i.e., $\sigma _{yx}=\frac{%
e^{2}}{h}%
\mathbb{Z}
$, where $%
\mathbb{Z}
$ is in the domain of integers or the first Chern numbers. In doing the
integration with respect to time, $t$, we need to examine the implicit
time-dependence of the matrix element of $G^{<}$ in the 'pullback'
representation defined below. This means reverting to the matrix
representation of Eq. (\ref{hallcond}).

To prove that Eq. (\ref{hallcond}) gives $\sigma _{yx}=\frac{e^{2}}{h}n$,
where $n\in 
\mathbb{Z}
$, we need to transform the integral of the equation to the curvature of the
Berry connection in a closed loop. This necessitates a 'pullback' (i.e.,
reverting to matrix elements) of Eq. (\ref{hallcond}).

The pullback procedure is founded on the observation that the proper
phase-space integral of a product of the respective lattice Weyl transform
of two operators is equivalent to taking the trace of the product of the
same two operators in any choosen basis states of the system. The details of
the pullback procedure or inverse LWT is given in the Appendix B \cite{arxiv}

The result of converting the quantum transport equation in the transformed
space, Eq. (\ref{eqKE}), to the untransformed space by undoing or 'pulling
back' the lattice Weyl transformation $\mathcal{W}$, amounts to canceling $%
\mathcal{W}$ in both side of the equation given by, 
\begin{eqnarray}
&&\mathcal{W}\left\{ \left\langle \hat{\jmath}_{y}\left( t\right)
\right\rangle \right\}  \notag \\
&=&\mathcal{W}\left\{ 
\begin{array}{c}
\frac{e^{2}}{h}\left\vert \vec{F}\right\vert \frac{1}{\left( 2\pi \hbar
\right) }\int \int \int d\mathcal{\vec{K}}_{x}d\mathcal{\vec{K}}_{y}dt \\ 
\times \sum\limits_{\alpha ,\beta }\left[ 
\begin{array}{c}
\left( E_{\beta }\left( \mathcal{\vec{K}},\mathcal{E}\right) -E_{\alpha
}\left( \mathcal{\vec{K}},\mathcal{E}\right) \right) \\ 
\times \left\{ 
\begin{array}{c}
\left\langle \alpha ,\frac{\partial }{\partial \mathcal{\vec{K}}_{x}}%
\mathcal{\vec{K}},\mathcal{E}\right\vert \left\vert \beta ,\mathcal{\vec{K}},%
\mathcal{E}\right\rangle \left\langle \beta ,\mathcal{\vec{K}},\mathcal{E}%
\right\vert \left\vert \alpha ,\frac{\partial }{\partial \mathcal{\vec{K}}%
_{y}}\mathcal{\vec{K}},\mathcal{E}\right\rangle \\ 
-\left\langle \alpha ,\frac{\partial }{\partial \mathcal{\vec{K}}_{y}}%
\mathcal{\vec{K}},\mathcal{E}\right\vert \left\vert \beta ,\mathcal{\vec{K}},%
\mathcal{E}\right\rangle \left\langle \beta ,\mathcal{\vec{K}},\mathcal{E}%
\right\vert \left\vert \alpha ,\frac{\partial }{\partial \mathcal{\vec{K}}%
_{x}}\mathcal{\vec{K}},\mathcal{E}\right\rangle%
\end{array}%
\right\} \\ 
\times \ e^{i\omega _{\alpha \beta }t}f\left( E_{\alpha }\right)%
\end{array}%
\right]%
\end{array}%
\right\} ,  \notag \\
&&  \label{pullback1}
\end{eqnarray}%
where%
\begin{equation*}
\mathcal{W}\left\{ \left\langle \hat{\jmath}_{y}\left( t\right)
\right\rangle \right\} =\left( \frac{a}{\left( 2\pi \hbar \right) }\right)
^{2}\int d\mathcal{\vec{K}}_{x}d\mathcal{\vec{K}}_{y}\frac{e}{a^{2}}\frac{%
\partial \mathcal{E}}{\partial \mathcal{\vec{K}}_{y}}\left[ -iG^{<}\left( 
\mathcal{\vec{K}},\mathcal{E}\right) \right] \text{.}
\end{equation*}

The time integral of the RHS amounts to taking zero-order time dependence
[zero electric field] of the rest of the integrand, then we have for the
remaining time-dependence, explicitly integrated as, 
\begin{equation*}
\int\limits_{-\infty }^{0}dt\exp i\omega _{\alpha \beta }t=\left. \frac{\exp
i\omega _{\alpha \beta }t}{i\omega _{\alpha \beta }}\right\vert _{\tau
=-\infty }^{\tau =0}=\frac{1}{i\omega _{\alpha \beta }}.
\end{equation*}%
Thus eliminating the time integral we finally obtain. 
\begin{eqnarray}
&&\left\langle \hat{\jmath}_{y}\left( t\right) \right\rangle  \notag \\
&=&-i\frac{e^{2}}{h}\left\vert \vec{F}\right\vert \frac{1}{\left( 2\pi \hbar
\right) }\int \int d\mathcal{\vec{K}}_{x}d\mathcal{\vec{K}}%
_{y}\sum\limits_{\alpha ,\beta }\left[ 
\begin{array}{c}
f\left( E_{\alpha }\right) \left( -\frac{\hbar \omega _{\alpha \beta }}{%
\omega _{\alpha \beta }}\right) \\ 
\times \left\{ 
\begin{array}{c}
\left\langle \alpha ,\frac{\partial }{\partial \mathcal{\vec{K}}_{x}}%
\mathcal{\vec{K}},\mathcal{E}\right\vert \left\vert \beta ,\mathcal{\vec{K}},%
\mathcal{E}\right\rangle \left\langle \beta ,\mathcal{\vec{K}},\mathcal{E}%
\right\vert \left\vert \alpha ,\frac{\partial }{\partial \mathcal{\vec{K}}%
_{y}}\mathcal{\vec{K}},\mathcal{E}\right\rangle \\ 
-\left\langle \alpha ,\frac{\partial }{\partial \mathcal{\vec{K}}_{y}}%
\mathcal{\vec{K}},\mathcal{E}\right\vert \left\vert \beta ,\mathcal{\vec{K}},%
\mathcal{E}\right\rangle \left\langle \beta ,\mathcal{\vec{K}},\mathcal{E}%
\right\vert \left\vert \alpha ,\frac{\partial }{\partial \mathcal{\vec{K}}%
_{x}}\mathcal{\vec{K}},\mathcal{E}\right\rangle%
\end{array}%
\right\}%
\end{array}%
\right]  \notag \\
&&\left\langle \hat{\jmath}_{y}\left( t\right) \right\rangle  \notag \\
&=&i\frac{e^{2}}{h}\left\vert \vec{F}\right\vert \frac{1}{\left( 2\pi
\right) }\int \int dk_{x}dk_{y}\sum\limits_{\alpha ,\beta }f\left( E_{\alpha
}\right) \left\{ 
\begin{array}{c}
\left\langle \alpha ,\frac{\partial }{\partial k_{x}}\mathcal{\vec{K}},%
\mathcal{E}\right\vert \left\vert \beta ,\mathcal{\vec{K}},\mathcal{E}%
\right\rangle \left\langle \beta ,\mathcal{\vec{K}},\mathcal{E}\right\vert
\left\vert \alpha ,\frac{\partial }{\partial k_{y}}\mathcal{\vec{K}},%
\mathcal{E}\right\rangle \\ 
-\left\langle \alpha ,\frac{\partial }{\partial k_{y}}\mathcal{\vec{K}},%
\mathcal{E}\right\vert \left\vert \beta ,\mathcal{\vec{K}},\mathcal{E}%
\right\rangle \left\langle \beta ,\mathcal{\vec{K}},\mathcal{E}\right\vert
\left\vert \alpha ,\frac{\partial }{\partial k_{x}}\mathcal{\vec{K}},%
\mathcal{E}\right\rangle%
\end{array}%
\right\}  \label{notint}
\end{eqnarray}%
Taking the Fourier transform of both sides, we obtain%
\begin{eqnarray}
&&\left\langle \hat{\jmath}_{y}\left( \omega \right) \right\rangle  \notag \\
&=&i\frac{e^{2}}{h}\left\vert \vec{F}\right\vert \frac{\delta \left( \omega
\right) }{\left( 2\pi \right) }\int \int dk_{x}dk_{y}\sum\limits_{\alpha
,\beta }\left\{ 
\begin{array}{c}
\left\langle \alpha ,\frac{\partial }{\partial k_{x}}\mathcal{\vec{K}},%
\mathcal{E}\right\vert \left\vert \beta ,\mathcal{\vec{K}},\mathcal{E}%
\right\rangle \left\langle \beta ,\mathcal{\vec{K}},\mathcal{E}\right\vert
\left\vert \alpha ,\frac{\partial }{\partial k_{y}}\mathcal{\vec{K}},%
\mathcal{E}\right\rangle \\ 
-\left\langle \alpha ,\frac{\partial }{\partial k_{y}}\mathcal{\vec{K}},%
\mathcal{E}\right\vert \left\vert \beta ,\mathcal{\vec{K}},\mathcal{E}%
\right\rangle \left\langle \beta ,\mathcal{\vec{K}},\mathcal{E}\right\vert
\left\vert \alpha ,\frac{\partial }{\partial k_{x}}\mathcal{\vec{K}},%
\mathcal{E}\right\rangle%
\end{array}%
\right\}  \notag \\
&&\times f\left( E_{\alpha }\right) .  \label{currenteq}
\end{eqnarray}%
Taking the limit $\omega \Longrightarrow 0$ and summing over the states $%
\beta $, we readily obtain the conductivity, $\sigma _{yx}$. 
\begin{equation}
\sigma _{yx}=\frac{e^{2}}{h}\sum\limits_{\alpha }f\left( E_{\alpha }\right) 
\frac{i}{\left( 2\pi \right) }\int \int dk_{x}dk_{y}\left[ 
\begin{array}{c}
\left\langle \alpha ,\frac{\partial }{\partial k_{x}}\mathcal{\vec{K}},%
\mathcal{E}\right\vert \left\vert \alpha ,\frac{\partial }{\partial k_{y}}%
\mathcal{\vec{K}},\mathcal{E}\right\rangle \\ 
-\left\langle \alpha ,\frac{\partial }{\partial k_{y}}\mathcal{\vec{K}},%
\mathcal{E}\right\vert \left\vert \alpha ,\frac{\partial }{\partial k_{x}}%
\mathcal{\vec{K}},\mathcal{E}\right\rangle%
\end{array}%
\right] \text{,}  \label{hall}
\end{equation}%
where $\mathcal{\vec{K}\Longrightarrow }\vec{p}=\hbar \vec{k}$ from Eq. (\ref%
{kappa}), with Jacobian unity, in both Eqs. (\ref{currenteq}) and (\ref{hall}%
). This is the same expression that can be obtained to derive the integer
quantum Hall effect from Kubo formula \cite{tknn}.

We now prove that for each statevector, $\left\vert \alpha ,\vec{k}%
\right\rangle $, the expression, 
\begin{equation}
\frac{i}{\left( 2\pi \right) }\int \int dk_{x}dk_{y}\ f\left( E_{\alpha
}\left( \vec{k}\right) \right) \left[ \left\langle \frac{\partial }{\partial
k_{x}}\alpha ,\vec{k}\right\vert \frac{\partial }{\partial k_{y}}\left\vert
\alpha ,\vec{k}\right\rangle -\left\langle \frac{\partial }{\partial k_{y}}%
\alpha ,\vec{k}\right\vert \frac{\partial }{\partial k_{x}}\left\vert \alpha
,\vec{k}\right\rangle \right] \text{,}  \label{chernNum}
\end{equation}%
is the winding number around the contour of occupied energy-bands in the
Brillouin zone. First we can rewrite the terms within the square bracket as%
\begin{eqnarray}
&&\left[ \left\langle \frac{\partial }{\partial k_{x}}\alpha ,\vec{k}%
\right\vert \frac{\partial }{\partial k_{y}}\left\vert \alpha ,\vec{k}%
\right\rangle -\left\langle \frac{\partial }{\partial k_{y}}\alpha ,\vec{k}%
\right\vert \frac{\partial }{\partial k_{x}}\left\vert \alpha ,\vec{k}%
\right\rangle \right]  \notag \\
&=&\left\langle \frac{\partial }{\partial \vec{k}}\alpha ,\vec{k}\right\vert
\times \frac{\partial }{\partial \vec{k}}\left\vert \alpha ,\vec{k}%
\right\rangle =\nabla _{\vec{k}}\times \left\langle \alpha ,\vec{k}%
\right\vert \frac{\partial }{\partial \vec{k}}\left\vert \alpha ,\vec{k}%
\right\rangle \text{.}  \label{curlA}
\end{eqnarray}%
The last term indicates the operation of the curl of the Berry connection
which is related to the quantization of Hall conductivity. This quantization
is due to the uniqueness of the parallel-transported wavefunction. To
understand this, we refer the readers to the Appendix A which discusses how
the phase of the wavefunction relates to the Berry connection and Berry
curvature.

At low temperature, we can just write,%
\begin{eqnarray*}
\ \sigma _{yx} &=&\frac{ie^{2}}{2\pi \hbar }\frac{1}{\left( 2\pi \right) }%
\sum\limits_{\alpha }\int \int_{occupiedBZ}dk_{x}dk_{y}\ \left[ \nabla _{%
\vec{k}}\times \left\langle \alpha ,\vec{k}\right\vert \frac{\partial }{%
\partial \vec{k}}\left\vert \alpha ,\vec{k}\right\rangle \right] _{plane} \\
&=&\left( \frac{e^{2}}{h}\right) \frac{i}{\left( 2\pi \right) }%
\sum\limits_{\alpha }\oint dk_{c}\ \left[ \left\langle \alpha ,\vec{k}%
\right\vert \frac{\partial }{\partial k_{c}}\left\vert \alpha ,\vec{k}%
\right\rangle \right] _{contour}\text{.}
\end{eqnarray*}%
Now in parallel transport, 
\begin{eqnarray*}
\frac{\Delta \phi _{total}}{2\pi } &=&\frac{i}{2\pi }\oint dk_{c}\ \left[
\left\langle \alpha ,\vec{k}\right\vert \frac{\partial }{\partial k_{c}}%
\left\vert \alpha ,\vec{k}\right\rangle \right] _{contour}\text{ } \\
&=&n,
\end{eqnarray*}%
where $\frac{\Delta \phi _{total}}{2\pi }$ is the winding number or the
Chern number. Therefore,%
\begin{equation}
\ \sigma _{yx}=\ \frac{e^{2}}{h}\sum\limits_{\alpha }\frac{\Delta \phi
_{total}}{2\pi }=\sum\limits_{\alpha }\frac{e^{2}}{h}n_{\alpha }\text{,}
\label{QHE}
\end{equation}%
over all occupied bands $\alpha ,$ where $n_{\alpha }\in 
\mathbb{Z}
$ is the topological first Chern (or \textit{winding}) number. Thus the the
Hall conductivity is quantized in units of $\frac{e^{2}}{h}$, as derive from
Eq. (\ref{hallcond}) of the new quantum transport approach used here.

The above analysis generalizes the Kubo current-current formula (KCCF) used
by TKNN \cite{tknn}. In fact, KCCF can be derived from our quantum transport
equation, shown in Appendix.

\section*{B-S quantization of orbital magnetic moment,edge states}

In what follows, we will make use of the following identities, 
\begin{equation}
\left\langle \alpha ,\vec{p}\right\vert \vec{v}\left\vert \beta ,\vec{p}%
\right\rangle =\left( E_{\alpha }\left( \vec{p}\right) -E_{\beta }\left( 
\vec{p}\right) \right) \left\langle \alpha ,\vec{p}\right\vert \left( \nabla
_{\vec{p}}\right) \left\vert \beta ,\vec{p}\right\rangle  \label{jd1}
\end{equation}%
together with 
\begin{equation}
\mathbf{\hat{Q}}\left\vert \alpha ,\vec{p}\right\rangle =-i\hbar \nabla _{%
\vec{p}}\left\vert \alpha ,\vec{p}\right\rangle =-i\nabla _{k}\left\vert
\alpha ,\vec{p}\right\rangle  \label{qonp}
\end{equation}

The orbital magnetic moment is given by,%
\begin{eqnarray*}
\left\langle \vec{M}\right\rangle &=&Tr\rho \frac{e}{2m}\vec{L}=Tr\rho \frac{%
e}{2m}\vec{Q}\times \vec{P} \\
&=&-i\frac{e}{2}\sum\limits_{\alpha ,p;\beta ,p}f\left( E_{\alpha }\left( 
\vec{p}\right) \right) \left\langle \alpha ,\vec{p}\right\vert \nabla _{\vec{%
k}}\left\vert \beta ,\vec{p}\right\rangle \times \left\langle \beta ,\vec{p}%
\right\vert \vec{v}\left\vert \alpha ,\vec{p}\right\rangle .
\end{eqnarray*}%
Using Eq. (\ref{transv2}), we obtain%
\begin{equation*}
\left\langle \vec{M}\right\rangle =-i\frac{e}{2}\sum\limits_{\alpha ,p;\beta
,p}f\left( E_{\alpha }\left( \vec{p}\right) \right) \omega _{\beta \alpha
}\left\langle \alpha ,\vec{p}\right\vert \nabla _{\vec{k}}\left\vert \beta ,%
\vec{p}\right\rangle \times \left\langle \beta ,\vec{p}\right\vert \nabla _{%
\vec{k}}\left\vert \alpha ,\vec{p}\right\rangle \text{,}
\end{equation*}%
or%
\begin{equation*}
\left\langle \vec{M}\right\rangle =-i\frac{e}{2}\sum\limits_{\alpha ,p;\beta
,p}\frac{f\left( E_{\alpha }\left( \vec{p}\right) \right) }{\omega _{\alpha
\beta }}\left\langle \alpha ,\vec{p}\right\vert \vec{v}\left\vert \beta ,%
\vec{p}\right\rangle \times \left\langle \beta ,\vec{p}\right\vert \vec{v}%
\left\vert \alpha ,\vec{p}\right\rangle \text{.}
\end{equation*}%
So Berry's curvature implies the presence of orbital magnetic moment. \
Using Eq. (\ref{transv2}), we obtain as written in some literature%
\begin{eqnarray*}
\left\langle \vec{M}\right\rangle &=&-\frac{e}{2}\left( i\right) ^{2}\frac{%
\hbar ^{2}}{m^{2}}\sum\limits_{\alpha ,p;\beta ,p}f\left( E_{\alpha }\left( 
\vec{p}\right) \right) \frac{\left( E_{\alpha }\left( \vec{p}\right)
-E_{\beta }\left( \vec{p}\right) \right) }{i\hbar }\frac{\left\langle \alpha
,\vec{p}\right\vert m\vec{v}\left\vert \beta ,\vec{p}\right\rangle }{\left(
E_{\alpha }\left( \vec{p}\right) -E_{\beta }\left( \vec{p}\right) \right) }%
\times \frac{\left\langle \beta ,\vec{p}\right\vert m\vec{v}\left\vert
\alpha ,\vec{p}\right\rangle }{\left( E_{\alpha }\left( \vec{p}\right)
-E_{\beta }\left( \vec{p}\right) \right) } \\
&=&-i\frac{e\hbar }{2m^{2}}\sum\limits_{\alpha ,p;\beta ,p}f\left( E_{\alpha
}\left( \vec{p}\right) \right) \frac{\left\langle \alpha ,\vec{p}\right\vert
m\vec{v}\left\vert \beta ,\vec{p}\right\rangle \times \left\langle \beta ,%
\vec{p}\right\vert m\vec{v}\left\vert \alpha ,\vec{p}\right\rangle }{\left(
E_{\alpha }\left( \vec{p}\right) -E_{\beta }\left( \vec{p}\right) \right) }%
\text{.}
\end{eqnarray*}

In fact we can extract the Berry curvature by rewriting in the Heisenberg
picture,%
\begin{eqnarray*}
\left\langle \vec{M}\right\rangle &=&Tr\rho \frac{e}{2m}L=Tr\rho \frac{e}{2m}%
\vec{Q}\times \vec{P} \\
&=&\frac{e}{2m}\sum\limits_{\alpha ,p;\beta ,p}f\left( E_{\alpha }\left( 
\vec{p}\right) \right) \left\langle \alpha ,\vec{p}\right\vert \vec{Q}%
\left\vert \beta ,\vec{p}\right\rangle \times \left\langle \beta ,\vec{p}%
\right\vert \vec{P}\left\vert \alpha ,\vec{p}\right\rangle \\
&=&\frac{e}{2}\sum\limits_{\alpha ,p;\beta ,p}f\left( E_{\alpha }\left( \vec{%
p}\right) \right) \left\langle \alpha ,\vec{p}\right\vert e^{\frac{i}{\hbar }%
Ht}\vec{Q}_{S}e^{-\frac{i}{\hbar }Ht}\left\vert \beta ,\vec{p}\right\rangle
\times \left\langle \beta ,\vec{p}\right\vert \vec{v}\left\vert \alpha ,\vec{%
p}\right\rangle \\
&=&\frac{e}{2}\sum\limits_{\alpha ,p;\beta ,p}f\left( E_{\alpha }\left( \vec{%
p}\right) \right) e^{i\omega _{\alpha \beta }t}\left\langle \alpha ,\vec{p}%
\right\vert \vec{Q}_{S}\left\vert \beta ,\vec{p}\right\rangle \times
\left\langle \beta ,\vec{p}\right\vert \vec{v}\left\vert \alpha ,\vec{p}%
\right\rangle \text{.}
\end{eqnarray*}%
From Eq. (\ref{qonp}), we obtain%
\begin{eqnarray*}
\left\langle \vec{M}\right\rangle &=&-i\frac{e}{2}\sum\limits_{\alpha
,p;\beta ,p}f\left( E_{\alpha }\left( \vec{p}\right) \right) e^{i\omega
_{\alpha \beta }t}\left\langle \alpha ,\vec{p}\right\vert \nabla _{\vec{k}%
}\left\vert \beta ,\vec{p}\right\rangle \times \left\langle \beta ,\vec{p}%
\right\vert \vec{v}\left\vert \alpha ,\vec{p}\right\rangle \\
&=&-i\frac{e}{2}\sum\limits_{\alpha ,p;\beta ,p}f\left( E_{\alpha }\left( 
\vec{p}\right) \right) e^{i\omega _{\alpha \beta }t}\left\langle \alpha ,%
\vec{p}\right\vert \nabla _{\vec{k}}\left\vert \beta ,\vec{p}\right\rangle
\times \left\langle \beta ,\vec{p}\right\vert \nabla _{\vec{k}}\left\vert
\alpha ,\vec{p}\right\rangle \omega _{\beta \alpha }\text{.}
\end{eqnarray*}%
Integrating the RHS with respect to time, the result is in dimensional units
of an orbital magnetic moment multiplied by time denoted by $\left\langle 
\mathcal{\vec{M}}\right\rangle $, 
\begin{eqnarray*}
\left\langle \mathcal{\vec{M}}\right\rangle &=&-i\frac{e}{2}%
\sum\limits_{\alpha ,p;\beta ,p}f\left( E_{\alpha }\left( \vec{p}\right)
\right) \int\limits_{-\infty }^{0}e^{i\omega _{\alpha \beta }t}dt\ \omega
_{\beta \alpha }\left\langle \alpha ,\vec{p}\right\vert \nabla _{\vec{k}%
}\left\vert \beta ,\vec{p}\right\rangle \times \left\langle \beta ,\vec{p}%
\right\vert \nabla _{\vec{k}}\left\vert \alpha ,\vec{p}\right\rangle \\
&=&-i\frac{e}{2}\sum\limits_{\alpha ,p;\beta ,p}f\left( E_{\alpha }\left( 
\vec{p}\right) \right) \frac{\omega _{\beta \alpha }}{i\omega _{\alpha \beta
}}dt\ \left\langle \alpha ,\vec{p}\right\vert \nabla _{\vec{k}}\left\vert
\beta ,\vec{p}\right\rangle \times \left\langle \beta ,\vec{p}\right\vert
\nabla _{\vec{k}}\left\vert \alpha ,\vec{p}\right\rangle \\
&=&-\frac{e}{2}\sum\limits_{\alpha ,p;\beta ,p}f\left( E_{\alpha }\left( 
\vec{p}\right) \right) \ \left\langle \alpha ,\nabla _{\vec{k}}\vec{p}%
\right\vert \left\vert \beta ,\vec{p}\right\rangle \times \left\langle \beta
,\vec{p}\right\vert \nabla _{\vec{k}}\left\vert \alpha ,\vec{p}\right\rangle
\\
&=&-\frac{e}{2}\sum\limits_{\alpha ,p}f\left( E_{\alpha }\left( \vec{p}%
\right) \right) \ \left\langle \alpha ,\nabla _{\vec{k}}\vec{p}\right\vert
\times \left\vert \alpha ,\nabla _{\vec{k}}\vec{p}\right\rangle \text{.}
\end{eqnarray*}%
The last line is just%
\begin{eqnarray}
\left\langle \mathcal{\vec{M}}\right\rangle &=&-\frac{e}{2}\frac{a^{2}}{%
\left( 2\pi \right) ^{2}}\sum\limits_{\alpha }\int \int dk_{x}dk_{y}\
f\left( E_{\alpha }\left( \vec{k}\right) \right) \left[ \left\langle \frac{%
\partial }{\partial k_{x}}\alpha ,\vec{k}\right\vert \frac{\partial }{%
\partial k_{y}}\left\vert \alpha ,\vec{k}\right\rangle -\left\langle \frac{%
\partial }{\partial k_{y}}\alpha ,\vec{k}\right\vert \frac{\partial }{%
\partial k_{x}}\left\vert \alpha ,\vec{k}\right\rangle \right]  \notag \\
&=&-\frac{e}{2}\frac{a^{2}}{\left( 2\pi \right) ^{2}}\sum\limits_{\alpha }\
f\left( E_{\alpha }\left( \vec{k}\right) \right) \int \int dk_{x}dk_{y}\
\left( \nabla _{\vec{k}}\times \left\langle \alpha ,\vec{k}\right\vert \frac{%
\partial }{\partial \vec{k}}\left\vert \alpha ,\vec{k}\right\rangle \right) 
\text{,}  \label{equt}
\end{eqnarray}%
explicitly revealing the Berry curvature derived from the expression for the
orbital magnetic moment. Equation (\ref{equt}) has the similar dimensional
units of a Bohr magneton multiplied by time $t$. The resulting quantization
of orbital motion leads to edge states and integer quantum Hall effect under
uniform magnetic fields. The general analysis of edge states marks the works
of Laughlin \cite{laughlin}, and Halperin \cite{halperin}.

\section*{Concluding remarks}

The thesis of the paper is that if we simply cast the B-S quantization
condition as a U(1) gauge theory, like the gauge field of the topological
quantum field theory (TQFT) via the Chern-Simons gauge theory, or
specifically as in topological band theory (TBT) of condensed matter physics
in terms of Berry connection and curvature to make it self-consistent, then
all the quantization method in all the physical phenomena treated in this
paper are unified. We have demonstrated how it permeates and pervades the
whole of modern quantum physics, implicitly identifying the B-S condition as
the forebear of modern geometrical or topological quantum theory and even
the geometric quantization in mathematics \cite{sansoneto}. We have shown
its place in the theory of the Nobel prize winning discoveries of quantum
Hall effect, Kosterlitz-Thouless transition in two-dimensional systems,
Haldane phase, and quantized vortices in superfluids. Here, we also show how
the U(1) gauge theory naturally give us the quantized magnetic charge of
Dirac monopole. The B-S quantization also leads to the expression of B-S
quantization of the quantum field theory of harmonic oscillator and in
coherent state representation.

The B-S quantization leads to quantum conductance, quantum flux,
Landau-level degeneracies, discrete flux in multi-loop holonomic FQHE,
discrete vortex charge, discrete Dirac monopole charge, energy levels of
harmonic oscillators, etc., as a theory of Berry connection. The B-S
quantization is thus expected to continue to hold more prominent role in the
advances of quantum physics and geometry.

Our new real-time SFLWT-NEGF quantum kinetic transport offered a new and
entirely open system approach that straightforwardly lead to B-S
quantization of IQHE. We have identified topological invariant in $\left( 
\vec{p},\vec{q};E,t\right) \Longrightarrow \left( \mathcal{\vec{K}},{\LARGE %
\varepsilon }\right) -$phase space quantum transport given Eq. (\ref%
{hallcond}), an integral expression which give results in $%
\mathbb{Z}
$ manifold, the so-called first Chern numbers. Moreover, the conventional
linearity in the electrical field strength may not be a necessary and
sufficient condition to prove the integer QHE, but rather it is the
first-order gradient expansion in the real-time SFLWT-NEGF quantum transport
equation \cite{Schade}.

In the case where in addition to uniform electric field a uniform magnetic
field is present, basically the canonical crystal momentum $\mathcal{K}$ in
Eq. (\ref{currenteq}) will incorporate the magnetic field as well through
the magnetic vector potential \cite{bj}. Thus, the result immediately
transfers to that of free electron gas under intense magnetic fields, on
which the original beautiful experiment was performed \cite{kdp}. It also
appears that electron-electron interaction of filled Landau levels which
does not break the symmetry of Eqs. (\ref{eq2})-(\ref{eq3}) can be treated
in similar manner \cite{Giuliani}.

It is worthwhile to mention that the real-time formalism of SFLWT-NEGF
multi-spinor quantum transport equations are also able to predict various
entanglements leading to different topological phases of low-dimensional and
nanostructured gapped condensed matter systems \cite{physicab}.

\begin{acknowledgement}
One of the authors (F.A.B.) is grateful for the hospitality of the USC
Department of Physics, for the support of the Balik Scientist Program of the
Philippine Council for Industry, Energy and Emerging Technology Research and
Development of the Department of Science and Technology (PCIEERD-DOST), and
for their Infrastructure Development Program (IDP) funding grant to
establish the LCFMNN at USC.
\end{acknowledgement}

\appendix

\section{Wavefunction phase under parallel transport}

In the adiabatic parallel transport case, the phase of the wavefunction is
determined by (assuming energy band $\alpha $ is far remove from the other
bands),%
\begin{equation*}
\frac{\partial \psi _{\alpha }\left( \vec{k}\right) }{\partial t}=\left(
-\left\langle \alpha ,\vec{k}\right\vert \frac{\partial }{\partial \vec{k}}%
\left\vert \alpha ,\vec{k}\right\rangle \cdot \frac{d\vec{k}}{dt}\right)
\psi _{\alpha }\left( \vec{k}\right) \text{.}
\end{equation*}%
Thus,%
\begin{equation*}
d\phi =i\left\langle \alpha ,\vec{k}\right\vert \frac{\partial }{\partial 
\vec{k}}\left\vert \alpha ,\vec{k}\right\rangle \cdot d\vec{k}\text{,}
\end{equation*}%
where $id\phi $ is the change of phase of the wavefunction along a curve in $%
\vec{k}$-space (Brillouin zone). Around a closed curve the total change of
phase must be a multiple of $2\pi $, i.e., $\Delta \phi =2\pi n$ $\left(
n\in 
\mathbb{Z}
\right) $ for the wavefunction to return to its original state. This
generalizes to vortex solution $X$-$Y$ model of spin systems and
two-dimensional hydrodynamics, and two-dimensional Bose-Einstein condensate.
The contour integral of the vortex solution 
\begin{equation*}
\frac{1}{2\pi }\oint \nabla _{\vec{k}}\phi \cdot dl=m
\end{equation*}%
is the Bohr-Sommerfeld quantization condition, Eq. (\ref{eq1}), reminiscent
of the way Landau level degeneracy is calculated.

\section{'Pullback' procedure by reverting to matrix elements}

Consider the integrand in Eq. (\ref{hallcond}) given by the partial
derivatives of lattice Weyl transformed quantities.%
\begin{eqnarray}
&&\left[ \frac{\partial ^{\left( a\right) }}{\partial \mathcal{\vec{K}}_{x}}%
\frac{\partial ^{\left( b\right) }}{\partial \mathcal{\vec{K}}_{y}}-\frac{%
\partial ^{\left( a\right) }}{\partial \mathcal{\vec{K}}_{y}}\frac{\partial
^{\left( b\right) }}{\partial \mathcal{\vec{K}}_{x}}\right] H^{\left(
a\right) }\left( \mathcal{\vec{K}},\mathcal{E}\right) \left( -iG^{<\left(
b\right) }\left( \mathcal{\vec{K}},\mathcal{E}\right) \right)  \notag \\
&=&\left[ \frac{\partial H^{\left( a\right) }\left( \mathcal{\vec{K}},%
\mathcal{E}\right) }{\partial \mathcal{\vec{K}}_{x}}\frac{\partial
G^{<\left( b\right) }\left( \mathcal{\vec{K}},\mathcal{E}\right) }{\partial 
\mathcal{\vec{K}}_{y}}-\frac{\partial H^{\left( a\right) }\left( \mathcal{%
\vec{K}},\mathcal{E}\right) }{\partial \mathcal{\vec{K}}_{y}}\frac{\partial
G^{<\left( b\right) }\left( \mathcal{\vec{K}},\mathcal{E}\right) }{\partial 
\mathcal{\vec{K}}_{x}}\right] \text{.}  \label{kxky}
\end{eqnarray}%
From Eq. (\ref{sump}) this can be written as a lattice Weyl transform $%
\mathcal{W}$ in the form,%
\begin{eqnarray}
\frac{\partial H^{\left( a\right) }\left( \mathcal{\vec{K}},\mathcal{E}%
\right) }{\partial \mathcal{\vec{K}}_{x}} &=&\mathcal{W}\left\{ \left( \frac{%
\partial }{\partial \mathcal{\vec{K}}_{x}^{\alpha }}+\frac{\partial }{%
\partial \mathcal{\vec{K}}_{x}^{^{\beta }}}\right) \left\langle \alpha ,%
\mathcal{\vec{K}},\mathcal{E}\right\vert \hat{H}\left\vert \beta ,\mathcal{%
\vec{K}},\mathcal{E}\right\rangle \right\}  \notag \\
&=&\mathcal{W}\left\{ \left( E_{\beta }\left( \mathcal{\vec{K}},\mathcal{E}%
\right) -E_{\alpha }\left( \mathcal{\vec{K}},\mathcal{E}\right) \right)
\left\langle \alpha ,\frac{\partial }{\partial \mathcal{\vec{K}}_{x}}%
\mathcal{\vec{K}},\mathcal{E}\right\vert \left\vert \beta ,\mathcal{\vec{K}},%
\mathcal{E}\right\rangle \right\} \text{,}  \label{LWTdhdKx}
\end{eqnarray}%
where $\left\langle \alpha ,\frac{\partial }{\partial \mathcal{\vec{K}}_{x}}%
\mathcal{\vec{K}},\mathcal{E}\right\vert $ symbolically denotes derivative
with respect to $\mathcal{\vec{K}}_{x}$ of the state vector $\left\langle
\alpha ,\mathcal{\vec{K}},\mathcal{E}\right\vert $ labeled by the three
quantum labels. Likewise for $\left\vert \beta ,\frac{\partial }{\partial 
\mathcal{\vec{K}}_{x}}\mathcal{\vec{K}},\mathcal{E}\right\rangle $ .

We also have%
\begin{equation*}
\frac{\partial G^{<\left( b\right) }\left( \mathcal{\vec{K}},\mathcal{E}%
\right) }{\partial \mathcal{K}_{y}}=\mathcal{W}\left\{ \left( \frac{\partial 
}{\partial \mathcal{\vec{K}}_{y}^{\beta }}+\frac{\partial }{\partial 
\mathcal{\vec{K}}_{y}^{^{\alpha }}}\right) \left\langle \beta ,\mathcal{\vec{%
K}},\mathcal{E}\right\vert \left( i\hat{\rho}\right) \left\vert \alpha ,%
\mathcal{\vec{K}},\mathcal{E}\right\rangle \right\} \text{,}
\end{equation*}%
where $\hat{\rho}$ is the density matrix operator. From Eq. (\ref{timedepend}%
), we take the time dependence of $\left\langle \beta ,\mathcal{\vec{K}},%
\mathcal{E}\right\vert \left( i\hat{\rho}\right) \left\vert \alpha ,\mathcal{%
\vec{K}},\mathcal{E}\right\rangle $ to be given by $i\left\langle \beta ,%
\mathcal{\vec{K}},\mathcal{E}\right\vert \hat{\rho}\left( 0\right)
\left\vert \alpha ,\mathcal{\vec{K}},\mathcal{E}\right\rangle e^{i\omega
_{\alpha \beta }t}$.

We have\footnote{%
Here we use the definition of Green's function without the factor $\hbar $,
following traditional treatments, i.e. $\rho \left( 1,2\right)
=-iG^{<}\left( 1,2\right) $.}%
\begin{equation*}
\frac{\partial G^{<\left( b\right) }\left( \mathcal{\vec{K}},\mathcal{E}%
\right) }{\partial \mathcal{K}_{y}}=\mathcal{W}\left\{ 
\begin{array}{c}
\left\langle \beta ,\frac{\partial }{\partial \mathcal{\vec{K}}_{y}}\mathcal{%
\vec{K}},\mathcal{E}\right\vert \left( i\hat{\rho}_{0}\right) \left\vert
\alpha ,\mathcal{\vec{K}},\mathcal{E}\right\rangle \\ 
+\left\langle \beta ,\mathcal{\vec{K}},\mathcal{E}\right\vert i\hat{\rho}%
_{0}\left\vert \alpha ,\frac{\partial }{\partial \mathcal{\vec{K}}_{y}}%
\mathcal{\vec{K}},\mathcal{E}\right\rangle%
\end{array}%
\right\} e^{i\omega _{\alpha \beta }t}\text{.}
\end{equation*}%
The density matrix operator $\hat{\rho}_{0}$\ is of the form,%
\begin{equation*}
\left\langle m\right\vert \hat{\rho}_{o}\left\vert n\right\rangle =f\left(
E_{n}\right) \delta _{mn}\text{ or }f\left( E_{m}\right) \delta _{mn}
\end{equation*}%
where the weight function is the Fermi-Dirac function. Hence%
\begin{eqnarray*}
i\hat{\rho}_{o}\left\vert \alpha ,\mathcal{\vec{K}},\mathcal{E}\right\rangle
&=&i\sum\limits_{\gamma }\left\vert \gamma ,\mathcal{\vec{K}},\mathcal{E}%
\right\rangle \rho _{0}^{\gamma }\left\langle \gamma ,\mathcal{\vec{K}},%
\mathcal{E}\right\vert \left\vert \alpha ,\mathcal{\vec{K}},\mathcal{E}%
\right\rangle \\
&=&i\left\vert \alpha ,\mathcal{\vec{K}},\mathcal{E}\right\rangle f\left(
E_{\alpha }\right) \text{.}
\end{eqnarray*}%
Similarly,%
\begin{eqnarray*}
i\left\langle \beta ,\mathcal{\vec{K}},\mathcal{E}\right\vert \left( \hat{%
\rho}_{0}\right) &=&i\left\langle \beta ,\mathcal{\vec{K}},\mathcal{E}%
\right\vert \sum\limits_{\gamma }\left\vert \gamma ,\mathcal{\vec{K}},%
\mathcal{E}\right\rangle \rho _{0}\left\langle \gamma ,\mathcal{\vec{K}},%
\mathcal{E}\right\vert \\
&=&if\left( E_{\beta }\right) \left\langle \beta ,\mathcal{\vec{K}},\mathcal{%
E}\right\vert \text{.}
\end{eqnarray*}%
Hence%
\begin{equation*}
\frac{\partial G^{<\left( b\right) }\left( \mathcal{\vec{K}},\mathcal{E}%
\right) }{\partial \mathcal{K}_{y}}=\mathcal{W}\left\{ 
\begin{array}{c}
i\left\langle \beta ,\frac{\partial }{\partial \mathcal{\vec{K}}_{y}}%
\mathcal{\vec{K}},\mathcal{E}\right\vert \left\vert \alpha ,\mathcal{\vec{K}}%
,\mathcal{E}\right\rangle \rho _{0}^{\alpha } \\ 
i\left\langle \beta ,\mathcal{\vec{K}},\mathcal{E}\right\vert \left\vert
\alpha ,\frac{\partial }{\partial \mathcal{\vec{K}}_{y}}\mathcal{\vec{K}},%
\mathcal{E}\right\rangle \rho _{0}^{\beta }%
\end{array}%
\right\} e^{i\omega _{\alpha \beta }t}\text{.}
\end{equation*}%
Shifting the first derivative to the right, we have 
\begin{equation*}
\frac{\partial G^{<\left( b\right) }\left( \mathcal{\vec{K}},\mathcal{E}%
\right) }{\partial \mathcal{K}_{y}}=\mathcal{W}\left[ \left\{ i\left(
f\left( E_{\beta }\right) -f\left( E_{\alpha }\right) \right) \left\langle
\beta ,\mathcal{\vec{K}},\mathcal{E}\right\vert \left\vert \alpha ,\frac{%
\partial }{\partial \mathcal{\vec{K}}_{y}}\mathcal{\vec{K}},\mathcal{E}%
\right\rangle \right\} e^{i\omega _{\alpha \beta }t}\right] \text{.}
\end{equation*}%
For energy scale it is convenient to choose to use $f\left( E_{\alpha
}\right) $ in the above equation, with the viewpoint that $\alpha $-state is
far remove from the $\beta $-state in gapped states, so that we can set $%
f\left( E_{\beta }\right) \simeq 0$. The case $\alpha =\beta $ is
indeterminate so that by setting $f\left( E_{\beta }\right) \simeq 0$
renders the summation to be well-defined. Therefore%
\begin{eqnarray*}
&&\frac{\partial H\left( \mathcal{\vec{K}},\mathcal{E}\right) }{\partial 
\mathcal{K}_{x}}\frac{\partial G^{<}\left( \mathcal{\vec{K}},\mathcal{E}%
\right) }{\partial \mathcal{K}_{y}} \\
&=&\left\{ 
\begin{array}{c}
\mathcal{W}\left[ \left( E_{\beta }\left( \mathcal{\vec{K}},\mathcal{E}%
\right) -E_{\alpha }\left( \mathcal{\vec{K}},\mathcal{E}\right) \right)
\left\{ \left\langle \alpha ,\frac{\partial }{\partial \mathcal{\vec{K}}%
_{x}^{^{^{\prime \prime }}}}\mathcal{\vec{K}},\mathcal{E}\right\vert
\left\vert \beta ,\mathcal{\vec{K}},\mathcal{E}\right\rangle \right\} \right]
\\ 
\times \mathcal{W}\left[ \left\{ i\left\langle \beta ,\mathcal{\vec{K}},%
\mathcal{E}\right\vert \left\vert \alpha ,\frac{\partial }{\partial \mathcal{%
\vec{K}}_{y}}\mathcal{\vec{K}},\mathcal{E}\right\rangle \right\} \right]
f\left( E_{\alpha }\right) e^{i\omega _{\alpha \beta }t}%
\end{array}%
\right\} \text{.}
\end{eqnarray*}%
Since it appears as a product of two Weyl transforms, it must be a trace
formula in the \textit{untransformed or pulled back} version, i.e., for the
remaing indices $\alpha $ and $\beta $ we must be a summation, 
\begin{eqnarray*}
&&\frac{\partial H\left( \mathcal{\vec{K}},\mathcal{E}\right) }{\partial 
\mathcal{K}_{x}}\frac{\partial G^{<}\left( \mathcal{\vec{K}},\mathcal{E}%
\right) }{\partial \mathcal{K}_{y}} \\
&=&\mathcal{W}\left[ \left\{ 
\begin{array}{c}
\sum\limits_{\alpha ,\beta }\left( E_{\beta }\left( \mathcal{\vec{K}},%
\mathcal{E}\right) -E_{\alpha }\left( \mathcal{\vec{K}},\mathcal{E}\right)
\right) \\ 
\times \left\{ \left\langle \alpha ,\frac{\partial }{\partial \mathcal{K}_{x}%
}\mathcal{\vec{K}},\mathcal{E}\right\vert \left\vert \beta ,\mathcal{\vec{K}}%
,\mathcal{E}\right\rangle \right\} \left\{ \left\langle \beta ,\mathcal{\vec{%
K}},\mathcal{E}\right\vert \left\vert \alpha ,\frac{\partial }{\partial 
\mathcal{K}_{y}}\mathcal{\vec{K}},\mathcal{E}\right\rangle \right\}
e^{i\omega _{\alpha \beta }t}%
\end{array}%
\right\} i\left( f\left( E_{\alpha }\right) \right) \right] \text{.}
\end{eqnarray*}%
Similarly, we have%
\begin{eqnarray*}
&&\frac{\partial H\left( \mathcal{\vec{K}},\mathcal{E}\right) }{\partial 
\mathcal{K}_{y}}\frac{\partial G^{<}\left( \mathcal{\vec{K}},\mathcal{E}%
\right) }{\partial \mathcal{K}_{x}} \\
&=&\mathcal{W}\left[ \left\{ 
\begin{array}{c}
\sum\limits_{\alpha ,\beta }\left( E_{\beta }\left( \mathcal{\vec{K}},%
\mathcal{E}\right) -E_{\alpha }\left( \mathcal{\vec{K}},\mathcal{E}\right)
\right) \\ 
\times \left\{ \left\langle \alpha ,\frac{\partial }{\partial \mathcal{K}_{y}%
}\mathcal{\vec{K}},\mathcal{E}\right\vert \left\vert \beta ,\mathcal{\vec{K}}%
,\mathcal{E}\right\rangle \right\} \left\{ \left\langle \beta ,\mathcal{\vec{%
K}},\mathcal{E}\right\vert \left\vert \alpha ,\frac{\partial }{\partial 
\mathcal{K}_{x}}\mathcal{\vec{K}},\mathcal{E}\right\rangle \right\}
e^{i\omega _{\alpha \beta }t}%
\end{array}%
\right\} if\left( E_{\alpha }\right) \right] \text{.}
\end{eqnarray*}%
Therefore we obtain%
\begin{eqnarray*}
&&\left[ \frac{\partial H\left( \mathcal{\vec{K}},\mathcal{E}\right) }{%
\partial \mathcal{K}_{x}}\frac{\partial G^{<}\left( \mathcal{\vec{K}},%
\mathcal{E}\right) }{\partial \mathcal{K}_{y}}-\frac{\partial H\left( 
\mathcal{\vec{K}},\mathcal{E}\right) }{\partial \mathcal{K}_{y}}\frac{%
\partial G^{<}\left( \mathcal{\vec{K}},\mathcal{E}\right) }{\partial 
\mathcal{K}_{x}}\right] \\
&=&\mathcal{W}\left[ \left\{ 
\begin{array}{c}
\sum\limits_{\alpha ,\beta }\left( E_{\beta }\left( \mathcal{\vec{K}},%
\mathcal{E}\right) -E_{\alpha }\left( \mathcal{\vec{K}},\mathcal{E}\right)
\right) \\ 
\times \left[ 
\begin{array}{c}
\left\{ \left\langle \alpha ,\frac{\partial }{\partial \mathcal{K}_{x}}%
\mathcal{\vec{K}},\mathcal{E}\right\vert \left\vert \beta ,\mathcal{\vec{K}},%
\mathcal{E}\right\rangle \right\} \left\{ \left\langle \beta ,\mathcal{\vec{K%
}},\mathcal{E}\right\vert \left\vert \alpha ,\frac{\partial }{\partial 
\mathcal{K}_{y}}\mathcal{\vec{K}},\mathcal{E}\right\rangle \right\} \\ 
-\left\{ \left\langle \alpha ,\frac{\partial }{\partial \mathcal{K}_{y}}%
\mathcal{\vec{K}},\mathcal{E}\right\vert \left\vert \beta ,\mathcal{\vec{K}},%
\mathcal{E}\right\rangle \right\} \left\{ \left\langle \beta ,\mathcal{\vec{K%
}},\mathcal{E}\right\vert \left\vert \alpha ,\frac{\partial }{\partial 
\mathcal{K}_{x}}\mathcal{\vec{K}},\mathcal{E}\right\rangle \right\}%
\end{array}%
\right]%
\end{array}%
\right\} ie^{i\omega _{\alpha \beta }t}\left( f\left( E_{\alpha }\right)
\right) \right] \text{.}
\end{eqnarray*}%
Now the LHS of Eq. (\ref{eqKE}), namely 
\begin{equation}
\left( \frac{a}{\left( 2\pi \hbar \right) }\right) ^{2}\int d\mathcal{\vec{K}%
}_{x}d\mathcal{\vec{K}}_{y}\frac{e}{a^{2}}\frac{\partial \mathcal{E}}{%
\partial \mathcal{\vec{K}}_{y}}G^{<}\left( \mathcal{\vec{K}},\mathcal{E}%
\right) =\left( \frac{a}{\left( 2\pi \hbar \right) }\right) ^{2}\int d%
\mathcal{\vec{K}}_{x}d\mathcal{\vec{K}}_{y}\frac{e}{a^{2}}\frac{\partial
H_{0}}{\partial \mathcal{\vec{K}}_{y}}G^{<}\left( \mathcal{\vec{K}},\mathcal{%
E}\right) \text{.}  \label{LHS}
\end{equation}%
Using the result of Eq. (\ref{LWTdhdKx}), we have%
\begin{eqnarray*}
&&\frac{\partial H_{0}\left( \mathcal{\vec{K}},\mathcal{E}\right) }{\partial 
\mathcal{\vec{K}}_{y}} \\
&=&\mathcal{W}\left\{ \left[ \left( E_{\alpha }\left( \mathcal{\vec{K}},%
\mathcal{E}\right) -E_{\beta }\left( \mathcal{\vec{K}},\mathcal{E}\right)
\right) \right] \left\langle \alpha ,\mathcal{\vec{K}},\mathcal{E}%
\right\vert \left\vert \beta ,\frac{\partial }{\partial \mathcal{\vec{K}}%
_{y}^{^{^{\prime \prime }}}}\mathcal{\vec{K}},\mathcal{E}\right\rangle
\right\} \\
&=&\mathcal{W}\left\{ \omega _{\beta \alpha }\left\langle \alpha ,\frac{%
\partial }{\partial k_{y}}\mathcal{\vec{K}},\mathcal{E}\right\vert
\left\vert \beta ,\mathcal{\vec{K}},\mathcal{E}\right\rangle \right\}
=\left\langle \alpha ,\mathcal{\vec{K}},\mathcal{E}\right\vert
v_{y}\left\vert \beta ,\mathcal{\vec{K}},\mathcal{E}\right\rangle ,
\end{eqnarray*}%
where%
\begin{equation*}
\omega _{\beta \alpha }\left\langle \alpha ,\nabla _{\vec{k}}\vec{p}%
\right\vert \left\vert \beta ,\vec{p}\right\rangle =\left\langle \alpha ,%
\vec{p}\right\vert \vec{v}\left\vert \beta ,\vec{p}\right\rangle .
\end{equation*}%
Likewise%
\begin{equation*}
G^{<}\left( \mathcal{\vec{K}},\mathcal{E}\right) =i\mathcal{W}\left(
\left\langle \beta ,\mathcal{\vec{K}},\mathcal{E}\right\vert \hat{\rho}%
\left\vert \alpha ,\mathcal{\vec{K}},\mathcal{E}\right\rangle \right) .
\end{equation*}%
Again since Eq. (\ref{LHS}) is a product of lattice Weyl transform, it must
be a trace in the \textit{untransformed} version, i.e., 
\begin{eqnarray*}
&&\left( \frac{a}{\left( 2\pi \hbar \right) }\right) ^{2}\int d\mathcal{\vec{%
K}}_{x}d\mathcal{\vec{K}}_{y}\frac{e}{a^{2}}\frac{\partial H}{\partial 
\mathcal{\vec{K}}_{y}}G^{<}\left( \mathcal{\vec{K}},\mathcal{E}\right) \\
&=&\mathcal{W}\left\{ 
\begin{array}{c}
i\int \left( \frac{a}{\left( 2\pi \hbar \right) }\right) ^{2}d\mathcal{\vec{K%
}}_{x}d\mathcal{\vec{K}}_{y} \\ 
\times \sum\limits_{\alpha ,\beta }\left\langle \alpha ,\mathcal{\vec{K}},%
\mathcal{E}\right\vert \frac{e}{a^{2}}v_{y}\left\vert \beta ,\mathcal{\vec{K}%
},\mathcal{E}\right\rangle \left\langle \beta ,\mathcal{\vec{K}},\mathcal{E}%
\right\vert \hat{\rho}\left\vert \alpha ,\mathcal{\vec{K}},\mathcal{E}%
\right\rangle%
\end{array}%
\right\} \\
&=&\mathcal{W}\left\{ iTr\left( \frac{e}{a^{2}}\hat{v}_{y}\right) \hat{\rho}%
\right\} =i\mathcal{W}\left\{ Tr\left( \hat{\jmath}_{y}\hat{\rho}\right)
\right\} =i\mathcal{W}\left\{ Tr\left( \hat{\jmath}_{y}\ \hat{\rho}\right)
\right\} \\
&=&i\mathcal{W}\left\{ \left\langle \hat{\jmath}_{y}\left( t\right)
\right\rangle \right\} \text{.}
\end{eqnarray*}%
For calculating the conductivity, we are interested in the term multiplying
the first order in electric field. We can now convert the quantum transport
equation in the transformed space, Eq. (\ref{eqKE}), to the untransformed
space by undoing or 'pulling back' the lattice Weyl transformation $\mathcal{%
W}$, which amounts to canceling $\mathcal{W}$ in both side of the equation
given by, 
\begin{eqnarray}
&&\mathcal{W}\left\{ \left\langle \hat{\jmath}_{y}\left( t\right)
\right\rangle \right\}  \notag \\
&=&\mathcal{W}\left\{ 
\begin{array}{c}
\frac{e^{2}}{h}\left\vert \vec{F}\right\vert \frac{1}{\left( 2\pi \hbar
\right) }\int \int \int d\mathcal{\vec{K}}_{x}d\mathcal{\vec{K}}_{y}dt \\ 
\times \sum\limits_{\alpha ,\beta }\left[ 
\begin{array}{c}
\left( E_{\beta }\left( \mathcal{\vec{K}},\mathcal{E}\right) -E_{\alpha
}\left( \mathcal{\vec{K}},\mathcal{E}\right) \right) \\ 
\times \left\{ 
\begin{array}{c}
\left\langle \alpha ,\frac{\partial }{\partial \mathcal{\vec{K}}_{x}}%
\mathcal{\vec{K}},\mathcal{E}\right\vert \left\vert \beta ,\mathcal{\vec{K}},%
\mathcal{E}\right\rangle \left\langle \beta ,\mathcal{\vec{K}},\mathcal{E}%
\right\vert \left\vert \alpha ,\frac{\partial }{\partial \mathcal{\vec{K}}%
_{y}}\mathcal{\vec{K}},\mathcal{E}\right\rangle \\ 
-\left\langle \alpha ,\frac{\partial }{\partial \mathcal{\vec{K}}_{y}}%
\mathcal{\vec{K}},\mathcal{E}\right\vert \left\vert \beta ,\mathcal{\vec{K}},%
\mathcal{E}\right\rangle \left\langle \beta ,\mathcal{\vec{K}},\mathcal{E}%
\right\vert \left\vert \alpha ,\frac{\partial }{\partial \mathcal{\vec{K}}%
_{x}}\mathcal{\vec{K}},\mathcal{E}\right\rangle%
\end{array}%
\right\} e^{i\omega _{\alpha \beta }t}f\left( E_{\alpha }\right)%
\end{array}%
\right] \text{.}%
\end{array}%
\right\}  \notag \\
&&  \label{pullback}
\end{eqnarray}%
The time integral of the RHS amounts to taking zero-order time dependence
[zero electric field] of the rest of the integrand, then we have for the
remaining time-dependence, explicitly integrated as, 
\begin{eqnarray*}
\int\limits_{-\infty }^{0}dt\exp i\omega _{\alpha \beta }t &=&\left. \frac{%
\exp \exp i\omega _{\alpha \beta }t}{i\omega _{\alpha \beta }}\right\vert
_{\tau =-\infty }^{\tau =0} \\
&=&\left. \frac{\exp \left( i\left( \omega _{\alpha \beta }-i\eta \right)
\tau \right) }{i\omega _{\alpha \beta }}\right\vert _{\tau =-\infty }^{\tau
=0}=\frac{1}{i\omega _{\alpha \beta }}.
\end{eqnarray*}%
Thus eliminating the time integral we finally obtain, 
\begin{equation}
\left\langle \hat{\jmath}_{y}\left( t\right) \right\rangle =i\frac{e^{2}}{h}%
\left\vert \vec{F}\right\vert \frac{1}{\left( 2\pi \right) }\int \int
dk_{x}dk_{y}\sum\limits_{\alpha ,\beta }f\left( E_{\alpha }\right) \left\{ 
\begin{array}{c}
\left\langle \alpha ,\frac{\partial }{\partial k_{x}}\mathcal{\vec{K}},%
\mathcal{E}\right\vert \left\vert \beta ,\mathcal{\vec{K}},\mathcal{E}%
\right\rangle \left\langle \beta ,\mathcal{\vec{K}},\mathcal{E}\right\vert
\left\vert \alpha ,\frac{\partial }{\partial k_{y}}\mathcal{\vec{K}},%
\mathcal{E}\right\rangle \\ 
-\left\langle \alpha ,\frac{\partial }{\partial k_{y}}\mathcal{\vec{K}},%
\mathcal{E}\right\vert \left\vert \beta ,\mathcal{\vec{K}},\mathcal{E}%
\right\rangle \left\langle \beta ,\mathcal{\vec{K}},\mathcal{E}\right\vert
\left\vert \alpha ,\frac{\partial }{\partial k_{x}}\mathcal{\vec{K}},%
\mathcal{E}\right\rangle%
\end{array}%
\right\} .  \notag
\end{equation}%
Taking the Fourier transform of both sides, taking the limit $\omega
\Longrightarrow 0$ and summing over the states $\beta $, we readily obtain
the conductivity, $\sigma _{yx}$, 
\begin{equation}
\sigma _{yx}=\frac{e^{2}}{h}\sum\limits_{\alpha }f\left( E_{\alpha }\right) 
\frac{i}{\left( 2\pi \right) }\int \int dk_{x}dk_{y}\left[ 
\begin{array}{c}
\left\langle \alpha ,\frac{\partial }{\partial k_{x}}\mathcal{\vec{K}},%
\mathcal{E}\right\vert \left\vert \alpha ,\frac{\partial }{\partial k_{y}}%
\mathcal{\vec{K}},\mathcal{E}\right\rangle \\ 
-\left\langle \alpha ,\frac{\partial }{\partial k_{y}}\mathcal{\vec{K}},%
\mathcal{E}\right\vert \left\vert \alpha ,\frac{\partial }{\partial k_{x}}%
\mathcal{\vec{K}},\mathcal{E}\right\rangle%
\end{array}%
\right] \text{,}
\end{equation}%
where $\mathcal{\vec{K}\Longrightarrow }\vec{p}=\hbar \vec{k}$ from Eq. (\ref%
{kappa}), with Jacobian unity, in both Eqs. (\ref{currenteq}) and (\ref{hall}%
). This is the same expression that can be obtained from Kubo formula \cite%
{tknn}.

\section{Derivation of Kubo current-current correlation formula}

To touch base with a time-dependent perturbation of the Kubo current-current
correlation we recall that in this approach, a time varying electric field
is indirectly used. To get to QHE the limiting case of $\omega
\Longrightarrow 0$ is taken after Fourier transformation of a convolution
integral. In adapting to our approach, this means that the time integral in
the expression of the RHS of Eq. (\ref{pullback1}) when transformed to
current-current correlation is a convolution integral before taking the
Fourier transform.

We start with the RHS of Eq. (\ref{pullback1}), 
\begin{eqnarray*}
&&RHS \\
&=&\frac{e^{2}}{h}\frac{1}{\left( 2\pi \hbar \right) }\int \int \int d%
\mathcal{\vec{K}}_{x}d\mathcal{\vec{K}}_{y}dt\sum\limits_{\alpha ,\beta }%
\left[ 
\begin{array}{c}
\left( E_{\beta }\left( \mathcal{\vec{K}},\mathcal{E}\right) -E_{\alpha
}\left( \mathcal{\vec{K}},\mathcal{E}\right) \right) \\ 
\times \left\{ 
\begin{array}{c}
\left\langle \alpha ,\frac{\partial }{\hbar \partial k_{x}}\mathcal{\vec{K}},%
\mathcal{E}\right\vert \left\vert \beta ,\mathcal{\vec{K}},\mathcal{E}%
\right\rangle \left\langle \beta ,\mathcal{\vec{K}},\mathcal{E}\right\vert
\left\vert \alpha ,\frac{\partial }{\hbar \partial k_{y}}\mathcal{\vec{K}},%
\mathcal{E}\right\rangle \\ 
-\left\langle \alpha ,\frac{\partial }{\hbar \partial k_{y}}\mathcal{\vec{K}}%
,\mathcal{E}\right\vert \left\vert \beta ,\mathcal{\vec{K}},\mathcal{E}%
\right\rangle \left\langle \beta ,\mathcal{\vec{K}},\mathcal{E}\right\vert
\left\vert \alpha ,\frac{\partial }{\hbar \partial k_{x}}\mathcal{\vec{K}},%
\mathcal{E}\right\rangle%
\end{array}%
\right\} \\ 
\times f\left( E_{\alpha }\right) e^{i\left( \omega _{\alpha \beta }\right)
t}%
\end{array}%
\right] ,
\end{eqnarray*}%
\begin{eqnarray}
&&RHS  \notag \\
&=&\frac{e^{2}}{h}\frac{\hbar }{\left( 2\pi \hbar \right) \hbar ^{2}}\int
\int \int d\mathcal{\vec{K}}_{x}d\mathcal{\vec{K}}_{y}dt\sum\limits_{\alpha
,\beta }\left[ 
\begin{array}{c}
\omega _{\beta \alpha } \\ 
\times \left\{ 
\begin{array}{c}
\left\langle \alpha ,\frac{\partial }{\partial k_{x}}\mathcal{\vec{K}},%
\mathcal{E}\right\vert \left\vert \beta ,\mathcal{\vec{K}},\mathcal{E}%
\right\rangle \left\langle \beta ,\mathcal{\vec{K}},\mathcal{E}\right\vert
\left\vert \alpha ,\frac{\partial }{\partial k_{y}}\mathcal{\vec{K}},%
\mathcal{E}\right\rangle \\ 
-\left\langle \alpha ,\frac{\partial }{\partial k_{y}}\mathcal{\vec{K}},%
\mathcal{E}\right\vert \left\vert \beta ,\mathcal{\vec{K}},\mathcal{E}%
\right\rangle \left\langle \beta ,\mathcal{\vec{K}},\mathcal{E}\right\vert
\left\vert \alpha ,\frac{\partial }{\partial k_{x}}\mathcal{\vec{K}},%
\mathcal{E}\right\rangle%
\end{array}%
\right\} \\ 
\times f\left( E_{\alpha }\right) e^{i\left( \omega _{\alpha \beta }\right)
t}%
\end{array}%
\text{.}\right]  \notag \\
&&  \label{RHSpullback}
\end{eqnarray}%
We make use of the general relations%
\begin{equation}
\left\langle \alpha ,\nabla _{\vec{k}}\vec{p}\right\vert \left\vert \beta ,%
\vec{p}\right\rangle =\frac{\left\langle \alpha ,\vec{p}\right\vert \vec{v}%
\left\vert \beta ,\vec{p}\right\rangle }{\omega _{\beta \alpha }}
\label{transv1}
\end{equation}%
Similarly, we have%
\begin{equation}
\left\langle \beta ,\vec{p}\right\vert \nabla _{\vec{k}}\left\vert \alpha ,%
\vec{p}\right\rangle =\frac{\left\langle \beta ,\vec{p}\right\vert \vec{v}%
\left\vert \alpha ,\vec{p}\right\rangle }{\omega _{\beta \alpha }}\text{.}
\label{transv2}
\end{equation}%
Substituting in Eq. (\ref{RHSpullback}), we then have the convolution
integral with respect to time,%
\begin{eqnarray*}
&&RHS \\
&=&\frac{e^{2}}{h}\frac{1}{\left( 2\pi \hbar \right) \hbar }\int \int \int d%
\mathcal{\vec{K}}_{x}d\mathcal{\vec{K}}_{y}dt^{\prime } \\
&&\times \sum\limits_{\alpha ,\beta }\frac{1}{\omega _{\beta \alpha }}\left[ 
\begin{array}{c}
\left\{ 
\begin{array}{c}
\left\langle \alpha ,\mathcal{\vec{K}},\mathcal{E}\right\vert
v_{x}\left\vert \beta ,\mathcal{\vec{K}},\mathcal{E}\right\rangle
\left\langle \beta ,\mathcal{\vec{K}},\mathcal{E}\right\vert v_{y}\left(
t-t^{\prime }\right) \left\vert \alpha ,\mathcal{\vec{K}},\mathcal{E}%
\right\rangle \\ 
-\left\langle \alpha ,\frac{\partial }{\partial k_{y}}\mathcal{\vec{K}},%
\mathcal{E}\right\vert v_{y}\left( t-t^{\prime }\right) \left\vert \beta ,%
\mathcal{\vec{K}},\mathcal{E}\right\rangle \left\langle \beta ,\mathcal{\vec{%
K}},\mathcal{E}\right\vert v_{x}\left\vert \alpha ,\mathcal{\vec{K}},%
\mathcal{E}\right\rangle%
\end{array}%
\right\} \\ 
\times \ e^{i\left( \omega _{\alpha \beta }\right) t^{\prime }}f\left(
E_{\alpha }\right)%
\end{array}%
\right] \text{.}
\end{eqnarray*}%
Consider the following Fourier transformation, 
\begin{equation*}
\frac{1}{\sqrt{2\pi }}\int_{-\infty }^{\infty }e^{i\omega t}F\left( t\right)
dt=\frac{1}{\sqrt{2\pi }}\int_{-\infty }^{\infty }e^{i\omega
t}dt\int_{-\infty }^{0}f\left( t-t^{\prime }\right) g\left( t^{\prime
}\right) dt^{\prime }\text{,}
\end{equation*}%
\begin{equation*}
\frac{1}{\sqrt{2\pi }}\int_{-\infty }^{\infty }e^{i\omega t}F\left( t\right)
dt=\frac{1}{\sqrt{2\pi }}\int_{-\infty }^{\infty }e^{i\omega t^{\prime
}}g\left( t^{\prime }\right) dt^{\prime }\int_{-\infty }^{0}e^{i\omega
\alpha }f\left( \alpha \right) d\alpha \text{.}
\end{equation*}%
We can transform the range of integration as follows,%
\begin{eqnarray*}
\int_{-\infty }^{0}e^{i\omega \alpha }f\left( \alpha \right) d\alpha
&=&\int_{\infty }^{0}e^{-i\omega \alpha }f\left( -\alpha \right) \left(
-d\alpha \right) =\int_{0}^{\infty }e^{-i\omega \alpha }f\left( -\alpha
\right) d\alpha \\
&=&\int_{0}^{\infty }e^{-i\omega \alpha }f^{\dagger }\left( \alpha \right)
d\alpha =\int_{0}^{\infty }e^{-i\omega \alpha }f\left( \alpha \right)
d\alpha \text{,}
\end{eqnarray*}%
since $f\left( \alpha \right) =j\left( \alpha \right) $ is the observable
current density and hence self-adjoint. We can apply this result in what
follows.

Defining the current density as $j_{x}=\frac{ev_{x}}{a^{2}}$, we obtain,
after Fourier transforming the convolution integral as%
\begin{eqnarray}
&&RHS  \notag \\
&=&\frac{a^{2}}{\hbar \omega }\int \int \left( \frac{a}{\left( 2\pi \hbar
\right) }\right) ^{2}d\mathcal{\vec{K}}_{x}d\mathcal{\vec{K}}%
_{y}\int_{0}^{\infty }dt\sum\limits_{\alpha ,\beta }\left[ 
\begin{array}{c}
\left\{ 
\begin{array}{c}
\left\langle \alpha ,\mathcal{\vec{K}},\mathcal{E}\right\vert \frac{ev_{x}}{%
a^{2}}\left\vert \beta ,\mathcal{\vec{K}},\mathcal{E}\right\rangle
\left\langle \beta ,\mathcal{\vec{K}},\mathcal{E}\right\vert \frac{%
ev_{y}\left( t\right) }{a^{2}}\left\vert \alpha ,\mathcal{\vec{K}},\mathcal{E%
}\right\rangle \\ 
-\left\langle \alpha ,\mathcal{\vec{K}},\mathcal{E}\right\vert \frac{%
ev_{y}\left( t\right) }{a^{2}}\left\vert \beta ,\mathcal{\vec{K}},\mathcal{E}%
\right\rangle \left\langle \beta ,\mathcal{\vec{K}},\mathcal{E}\right\vert 
\frac{ev_{x}}{a^{2}}\left\vert \alpha ,\mathcal{\vec{K}},\mathcal{E}%
\right\rangle%
\end{array}%
\right\} \\ 
\times \ e^{-i\left( \omega -i\eta \right) t}f\left( E_{\alpha }\right)%
\end{array}%
\right]  \notag \\
&=&\frac{a^{2}}{\hbar \omega }\int_{0}^{\infty }dt\int \int \left( \frac{a}{%
\left( 2\pi \hbar \right) }\right) ^{2}d\mathcal{\vec{K}}_{x}d\mathcal{\vec{K%
}}_{y}\sum\limits_{\alpha ,\beta }\left[ 
\begin{array}{c}
\left\{ 
\begin{array}{c}
\left\langle \alpha ,\mathcal{\vec{K}},\mathcal{E}\right\vert j_{x}\left(
0\right) \left\vert \beta ,\mathcal{\vec{K}},\mathcal{E}\right\rangle
\left\langle \beta ,\mathcal{\vec{K}},\mathcal{E}\right\vert j_{y}\left(
t\right) \left\vert \alpha ,\mathcal{\vec{K}},\mathcal{E}\right\rangle \\ 
-\left\langle \alpha ,\mathcal{\vec{K}},\mathcal{E}\right\vert j_{y}\left(
t\right) \left\vert \beta ,\mathcal{\vec{K}},\mathcal{E}\right\rangle
\left\langle \beta ,\mathcal{\vec{K}},\mathcal{E}\right\vert j_{x}\left(
0\right) \left\vert \alpha ,\mathcal{\vec{K}},\mathcal{E}\right\rangle%
\end{array}%
\right\} \\ 
\times \ e^{-i\left( \omega -i\eta \right) t}f\left( E_{\alpha }\right)%
\end{array}%
\right]  \notag \\
&=&\frac{a^{2}}{\hbar \omega }\int_{0}^{\infty }dt\ Tr\rho _{0}\left\{ \left[
j_{x}\left( 0\right) ,j_{y}\left( t\right) \right] \right\} e^{-i\left(
\omega -i\eta \right) t}\text{,}  \label{kuboformula}
\end{eqnarray}%
where $\eta $ is just a regularization exponent at $\infty $. Therefore the
Kubo formula for the conductivity is given by%
\begin{equation}
\sigma _{yx}\left( t\right) =\frac{a^{2}}{\hbar \omega }\int_{0}^{\infty
}dt\ Tr\rho _{0}\left\{ \left[ j_{x}\left( 0\right) ,j_{y}\left( t\right) %
\right] \right\} e^{-i\left( \omega -i\eta \right) t}\text{.}
\label{kformula2}
\end{equation}
This is the Kubo current-current correlation formula for the Hall
conductivity. The way the B-S condition permeates the Kubo current-current
formula is implicit and was utilized by TKNN to derive the topological
anomalous (i.e., without magnetic field) IQHE.

\end{document}